%% file: bare_jrnl.tex
\documentclass[journal]{IEEEtran}
\ifCLASSINFOpdf
\else
\fi
\usepackage{acro}
\usepackage{graphicx}
\usepackage[font=small]{caption}
\usepackage{subcaption}
\usepackage{amsmath}
\usepackage{tabulary}
\usepackage{xcolor,colortbl}
\usepackage{multirow}

\definecolor{Gray}{gray}{0.75}
\definecolor{LightCyan}{rgb}{0.88,1,1}
\newcolumntype{a}{>{\columncolor{Gray}}c}
\newcolumntype{b}{>{\columncolor{white}}c}
\usepackage{cite}
\usepackage{tabularx}
\usepackage{booktabs}
\usepackage{bm}
\newcommand{\ra}[1]{\renewcommand{\arraystretch}{#1}}
\input{./acronym.tex}

\begin{document}
%
\title{Automatic Detection and Positioning of Ground Control Points Using TerraSAR-X Multi-Aspect Acquisitions}
%
%
%

\author{Sina~Montazeri,
        Christoph~Gisinger,
        Michael~Eineder,~\IEEEmembership{Fellow,~IEEE}
        and~Xiao Xiang~Zhu,~\IEEEmembership{Senior~Member,~IEEE,}

\thanks{Manuscript  received  June  29,  2017;  revised  September  8,  2017;  accepted October 12, 2017. This work was supported in part by the European Research Council through the European Union Horizon 2020 Research And Innovation Program   under   Grant   ERC-2016-StG-714087,   in   part   by  the   Helmholtz Association through the framework of the Young Investigators Group “SiPEO” under  Grant  VH-NG-1018,  and  in  part  by  Munich  Aerospace  e.V.  Fakultät für Luft- und Raumfahrt.
(Corresponding author: Xiao Xiang Zhu.)}
\thanks{S. Montazeri is with the Remote Sensing Technology Institute, German Aerospace Center, 82234 Wessling, Germany (e-mail: sina.montazeri@dlr.de). }
\thanks{C. Gisinger  is with the Remote Sensing Technology Institute, German Aerospace Center, 82234 Wessling, Germany.}
\thanks{M. Eineder is with the Remote Sensing Technology Institute, German Aerospace Center, 82234 Wessling, Germany, and also with the Chair of Remote Sensing Technology, Technische Universit{\"a}t M{\"u}nchen, 80333 Munich, Germany.}
\thanks{X.  X.  Zhu  is  with  the  Remote  Sensing  Technology  Institute,  German
Aerospace Center, 82234 Wessling, Germany, and also with the Signal Processing  for  Earth  Observation  (SiPEO),  Technische Universit{\"a}t M{\"u}nchen,  80333 Munich, Germany (e-mail: xiao.zhu@dlr.de).}

}
%
%

\markboth{IEEE Transactions on Geoscience and Remote Sensing, in press}%
{Shell \MakeLowercase{\textit{et al.}}: Bare Demo of IEEEtran.cls for IEEE Journals}
%



\maketitle

\begin{abstract}
\begingroup
    \fontsize{8pt}{8pt}\selectfont
         \textit{This is the pre-acceptance version, to read the final version please go to IEEE Transactions on Geoscience and Remote Sensing on IEEE XPlore. (DOI: 10.1109/TGRS.2017.2769078)}
\endgroup\\
Geodetic stereo Synthetic Aperture Radar (SAR) is capable of absolute three-dimensional localization of natural Persistent Scatterer (PS)s which allows for Ground Control Point (GCP) generation using only SAR data. The prerequisite for the method to achieve high precision results is the correct detection of common scatterers in SAR images acquired from different viewing geometries. In this contribution, we describe three strategies for automatic detection of identical targets in SAR images of urban areas taken from different orbit tracks. Moreover, a complete work-flow for automatic generation of large number of GCPs using SAR data is presented and its applicability is shown by exploiting TerraSAR-X (TS-X) high resolution spotlight images over the city of Oulu, Finland and a test site in Berlin, Germany.
\end{abstract}


%
\IEEEpeerreviewmaketitle

\section{Introduction}
\label{sec:intro}
%
%
%
%
\ac{sar} imaging geodesy and geodetic stereo \ac{sar} are relatively new techniques which aim at high precision absolute positioning of point targets in \ac{sar} images in \ac{2D} and \ac{3D}, respectively \cite{eineder_imaging_2011,cong_imaging_2012,gisinger_precise_2015}. The accuracy of both methods, when coupled with data from \ac{tsx} and TanDEM-X, is in the centimeter regime for targets with accurately known phase centers such as corner reflectors \cite{gisinger_precise_2015}. This level of accuracy is achievable due to the precise orbit determination \cite{hackel_model_2016} and instrument calibration of the aforementioned satellites followed by a thorough correction scheme which quantifies and removes the most prominent error sources affecting radar timing measurements. This paves the way for remotely sensed generation of \ac{gcp}s using only \ac{sar} data.\par

\begin{figure}
        \begin{subfigure}[t]{0.25\textwidth}
                \centering
                \includegraphics[width=.85\linewidth]{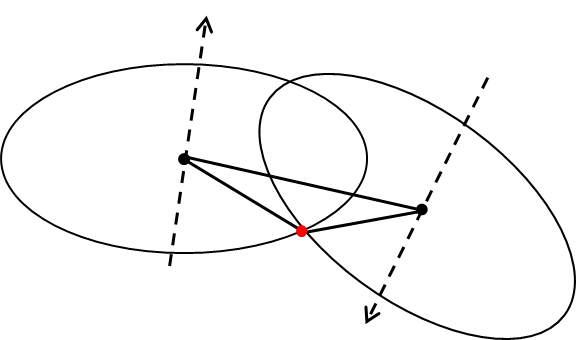}
                \caption{Cross-heading}
                \label{fig:cross}
        \end{subfigure}%
        \begin{subfigure}[t]{0.25\textwidth}
                \centering
                \includegraphics[width=.85\linewidth]{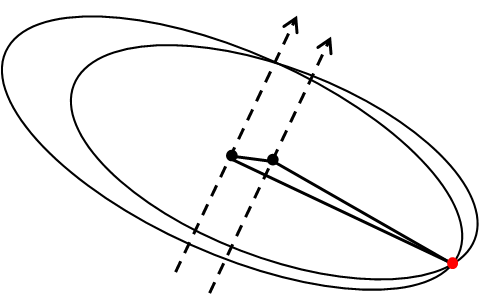}
                \caption{Same-heading}
                \label{fig:same}
        \end{subfigure}
        \caption{Localization of a point target (red dot) from (a) cross-heading and (b) same-heading satellite tracks. The satellites are shown by black dots; their trajectories are presented by dashed lines and the baselines are depicted by solid lines between the satellite positions. The black circles are defined by the range-Doppler equations and their intersection leads to the \ac{3D} position of the target.}\label{fig:Geo_config}
\end{figure}

The essential prerequisite for applying the geodetic stereo \ac{sar} method is the correct detection of identical scatterers in \ac{sar} images acquired from different geometries. In this regard, a target can be visible only from \textit{same-heading} orbits, i.e. exclusively ascending or descending orbits, or also from \textit{cross-heading} orbits, which include combinations of ascending and descending orbits. Conceptually, a target localized from the latter is favorable because of the more robust intersection geometry when compared to the former. This fact is demonstrated in Fig. \ref{fig:cross} where the intersection angle occurs at almost $90^{\circ}$ because of the large baseline between the satellites from cross-heading tracks. In Fig. \ref{fig:same}, the target is localized with satellites from same-heading tracks which force the baseline to be smaller and consequently the system of equations to solve for the \ac{3D} coordinates to be less sensitive for the perpendicular height direction. However, the rare occurrence of identical scatterers visible from cross-heading configurations as well as the challenging task of automatically detecting such targets, either from same- or cross-heading tracks, currently limit the applicability of geodetic stereo \ac{sar} for localization of large number of \ac{ps}s.

To overcome the limitation to some extent, this paper describes an automatic algorithm for detection and absolute positioning of large number of natural \ac{ps}s in \ac{sar} images of urban areas. The candidates are selected from both same-heading and cross-heading geometries based on methods relying on fusion of multitrack \ac{psi} point clouds, correspondence detection with optical data and utilizing vectorized road network data. The candidates are mainly chosen from the same-heading configuration because of the fact that for many \ac{ps}s the phase centers are assumed to remain unchanged in \ac{sar} images. On the other hand, additional candidates are chosen from cross-heading geometries, although in a small number, because conceptually they can be localized more precisely compared to the candidates from same-heading geometries. Coupled with the subsequent geodetic Stereo \ac{sar}, the proposed processing chain delivers sets of absolutely localized \ac{ps}s in an investigated area.\par
The remainder of the paper is organized as follows. Section \ref{sec:theory} reviews the theoretical background of the techniques utilized in this study and gives an overview of the recent advances and the motivation for this work. Section \ref{sec:detection} describes three methods for detecting identical \ac{ps}s visible in \ac{sar} images from same- and cross-heading tracks. In Section \ref{sec:chain}, the complete work-flow for generating high precision absolute \ac{gcp}s is explained. In section \ref{sec:results}, the applicability of the algorithm is demonstrated by exploiting \ac{tsx} high resolution spotlight images over the city of Oulu, Finland and a test site in Berlin, Germany, and finally the conclusions are drawn in section \ref{sec:conc}.

\section{High Precision Absolute \ac{2D} and \ac{3D} Positioning with \ac{tsx}}
\label{sec:theory}

At the core of high precision absolute positioning of candidate \ac{gcp}s using \ac{sar} data are the imaging geodesy and the stereo \ac{sar} methods. These methods are described in this section followed by the recent advances and applications which rely on absolute localization capability of \ac{tsx}. It is important to note that the complete explanation of the theory of the methods and their practical implementations are not in the scope of this paper. For full treatment of these topics, the interested reader is referred to \cite{eineder_imaging_2011,cong_imaging_2012,gisinger_precise_2015,hackel_model_2016,balss_high_2012,balss_precise_2014,balss_analysis_2014}.

\subsection{Background}
\label{ssec:backgr}

The \ac{sar} imaging geodesy technique aims at achieving \ac{2D} absolute pixel localization \cite{eineder_imaging_2011}. Based on the \ac{sar} measurement principle, a single pixel in a focused complex \ac{sar} image, processed to zero-Doppler coordinates, is characterized with two time tags: in the along-track direction, the time relative to the time of the closest approach defines the azimuth coordinate $t_{az}$ and in the across-track direction, the difference in the time travel of the transmitted and the received chirp at $t_{az}$ describes the range coordinate $\tau_{rg}$ \cite{cumming_digital_2005}. If we measure the radar timing coordinates ($t_{az}$, $\tau_{rg}$) for a point target located within the mentioned pixel, the following equations hold:

\begin{equation}
\label{eq:rg_zeroD}
\tau_{rg} = \frac{2 R}{c} + \delta\tau_{SD} + \delta\tau_{O} + \delta\tau_{F} + \delta\tau_{I} + \delta\tau_{T} + \delta\tau_{G},
\end{equation}
\begin{equation}\label{eq:az_zeroD}
t_{az} = t + \delta t_{SD} +  \delta t_{O} + \delta t_{F} + \delta t_{G},
\end{equation}

where $R$ is the geometric distance from the satellite to the center of the pixel in meters and $c$ is the speed of light in vacuum in $m/s$ while all the other terms are expressed in seconds; $t$ is the raw acquisition time, $\delta\tau_{SD}$ and $\delta t_{SD}$ are delays caused by satellite dynamics and electronics, $\delta\tau_{O}$ and $\delta t_{O}$ are the orbit inaccuracies, $\delta\tau_{F}$ and $\delta t_{F}$ are the feature localization error, $\delta\tau_{G}$ and $\delta t_{G}$ include the geodynamic effects all on range and azimuth timings, respectively while $\delta\tau_{I}$ and $\delta\tau_{T}$ are the ionospheric and the tropospheric delays considered only for range timings. The magnitude of the individual effects can be scaled to units of length by multiplying the range error terms with $\frac{c}{2}$ and the azimuth error terms with the platform's velocity. The outcomes vary from a couple of centimeters for the ionospheric effect, if the satellite operates in X-band, followed by decimeter regimes for satellite electronic delays and geodynamic effects for both components, to up to four meters for the tropospheric effect depending on the average incidence angle of the acquired TS-X images.\par
Imaging geodesy corrects for all the error terms in (\ref{eq:rg_zeroD}) and (\ref{eq:az_zeroD}) thus obtaining absolute range and azimuth timings. In this regard, the technique reduces the satellite dynamics effects by avoiding the stop-go approximation in the \ac{tmsp} and by taking into account the non-zero duration of the pulses and the internal delay caused by the instrument cables \cite{breit_terrasar-x_2004}. The propagation errors are estimated based on the path delays derived from the near-by \ac{gnss} stations or \ac{3D} integration through weather models followed by appropriate mapping functions \cite{cong_imaging_2012,gisinger_precise_2015,gisinger_atmospheric_2012}. For the geodynamic effects such as solid earth tides, plate tectonics, ocean loading and atmospheric loading, which change the position of a target on the ground, the corrections are applied based on models issued by the \ac{iers} \cite{petit_iers_2010}. Taking into account all the mentioned factors, \ac{sar} imaging geodesy is currently capable of providing range and azimuth measurements with 1.16 cm and 1.85 cm standard deviations, respectively \cite{dlr109261}.\par

If a target is visible in \ac{sar} images acquired from two or more different viewing geometries, then stereo \ac{sar} retrieves the \ac{3D} position of the target by combining the extracted timing information of the target from each \ac{sar} image. Furthermore, if the timing coordinates have been a-priori corrected for the error sources expressed in (\ref{eq:rg_zeroD}) and (\ref{eq:az_zeroD}), the method is called geodetic stereo \ac{sar} which allows for absolute \ac{3D} localization \cite{gisinger_precise_2015}. The relation between the \ac{2D} radar time coordinates of a specific target in the \ac{sar} image $\mathbf{x}_{T} = (t_{az},\tau_{rg})$ and its corresponding \ac{3D} coordinates on the ground $\mathbf{X}_T = ( X,Y,Z )$ is defined by the range-Doppler equation system \cite{cumming_digital_2005}:

\begin{equation}\label{eq:rg_eq}
|\textbf{X}_{S} - \textbf{X}_{T}| - c \cdot \tau_{rg} = 0,
\end{equation}
\begin{equation}
\label{eq:az_eq}
\frac{\dot{\mathbf{X}}_{S} (\mathbf{X}_{T} - \mathbf{X}_{S}) }{|\dot{\mathbf{X}}_{S}| |\textbf{X}_{S} - \textbf{X}_{T}|} = 0,
\end{equation}

with $\mathbf{X}_{S}$ and $\dot{\mathbf{X}}_{S}$ being the position and velocity vector of the satellite relative to $t_{az}$, and $\tau_{rg}$ being the calibrated two-way traveled time from the satellite to the target. $t_{az}$ is implicitly included in (\ref{eq:az_eq}) relating the state-vector of the satellite to the time of the acquisition via a polynomial model \cite{gisinger_precise_2015}. Equation (\ref{eq:rg_eq}) defines a sphere centered on $\mathbf{X}_{S}$ which reduces to a circle perpendicular to the satellite trajectory when coupled with the zero-Doppler plane described in (\ref{eq:az_eq}). Therefore $\mathbf{X}_{T}$ can be retrieved by including another set of timing observations from a different satellite position which evaluates the intersection point of the two circles, see Fig. \ref{fig:Geo_config}. The estimation of the coordinates is carried out by means of least squares plus stochastic modeling of range and azimuth using the variance component estimation (VCE) \cite{gisinger_precise_2015}. Precision of the estimated \ac{3D} coordinates depends on the \ac{scr} of the target, the precision of the external radar timing corrections, the separation in the viewing geometries and the number of acquisitions. Geodetic stereo \ac{sar} has been proven to be able to localize corner reflectors with \ac{3D} precision better than 4 cm and an absolute accuracy of 2-3 cm when compared to independently surveyed reference positions \cite{gisinger_precise_2015}.

\subsection{Recent advances and motivation}
\label{ssec:recent_advance}

In our previous research, geodetic stereo \ac{sar} has been also applied to small number of natural \ac{ps}s in urban areas where it could localize targets in \ac{3D} with a precision better than one decimeter using \ac{tsx} high resolution spotlight products \cite{gisinger_precise_2015}. The \ac{ps}s were manually extracted from \ac{sar} images and originated from building facades for candidates visible in same-heading tracks or from the base of street lights for candidates visible from cross-heading tracks.\par

In \cite{gisinger_absolute_2015}, the first attempt for automatic timing extraction and matching of limited number of \ac{ps}s originated from a building facade visible in \ac{tsx} images from two same-heading tracks was reported. In the study, the geodetic stereo \ac{sar} method was extended to include the secular movement of the \ac{ps}s as well as their \ac{3D} absolute positions. The averaged \ac{3D} precision was reported to be below one decimeter with encouraging results for estimating the plate tectonics using \ac{sar} data.\par

In \cite{zhu_geodetic_2016}, the concepts of imaging geodesy and stereo \ac{sar} were used to transform the relative estimates of \ac{sar} tomography into absolute \ac{3D} point clouds by absolutely localizing the manually extracted reference point. The method, termed geodetic \ac{sar} tomography, allows for generation of dense point clouds with an absolute localization accuracy in the order of 20 cm and is the basis for geodetic fusion of multi-aspect \ac{insar} point clouds. The latter enables the decomposition of deformation estimates from \ac{sar} tomography into highly detailed \ac{3D} displacement maps \cite{montazeri_three-dimensional_2016}.\par

Automatic extraction of \ac{gcp}s from \ac{sar} products have been carried out for \ac{tsx} and COSMO-SkyMed in \cite{balss_automated_2016} and \cite{notarnicola_automatic_2016}, respectively. Both methods focus on detection of stereo candidates that presumably originate from street lights or traffic signs and are visible in \ac{sar} images as bright isolated points. Therefore, the majority of \ac{ps}s in urban areas which stem from building corners or facades are not considered as potential candidates in these methods due to complex radar reflection properties in such scenarios. Furthermore, in \cite{balss_automated_2016} the \ac{3D} positioning is done only with same-heading geometry configurations and therefore the error ellipsoid of the scatterers' coordinates is highly skewed in the cross-range direction \cite{gisinger_precise_2015,dheenathayalan_high-precision_2016}. Nevertheless, the retrieved \ac{3D} coordinates of several candidates were compared to their true positions observed with \ac{gnss} which confirmed an absolute accuracy better than 20 cm in each coordinate component \cite{balss_automated_2016}.\par

Based on the above-mentioned studies, the motivation for carrying out this work is four-fold:

\begin{itemize}
\item The manual detection, extraction and matching of \ac{ps} candidates from \ac{sar} images acquired from different geometries is cumbersome and should be replaced by an automatic process.
\item The algorithm should be able to automatically detect and match identical \ac{ps}s visible from cross-heading geometries in order to boost the precision of the retrieved \ac{3D} coordinates.
\item The total number of high quality \ac{ps} candidates to be localized as \ac{gcp}s should be as large as possible. This indicates that the majority of \ac{ps}s in urban areas which stem from buildings should also be considered as candidates for \ac{3D} absolute localization from same-heading tracks.
\item The distribution of the \ac{gcp}s should be as homogeneous as possible in the entire investigated area. This allows for generation of an absolute reference network to be integrated into relative \ac{insar} techniques.
\end{itemize}

\section{Detection of Identical \ac{ps}s in Multi-Aspect \ac{sar} Images}
\label{sec:detection}

Detection of identical \ac{ps}s from \ac{sar} images acquired with different viewing geometries is a challenging task. This is because of the geometrical distortions of \ac{sar} images due to the oblique viewing geometry and less importantly the presence of speckle. Moreover, in urban areas captured by \ac{sar} sensors, which is the focus of this study, the backscattering mechanism is highly complex because of existence of several phase centers close to each other. Therefore, identical \ac{ps} matching becomes even more difficult for multi-aspect \ac{sar} images of urban areas.\par

In recent years, there have been several studies which explored the possibility to match features between \ac{sar} images. In \cite{suri_modifications_2010}, the capabilities of the conventional scale invariant feature transform (SIFT) algorithm \cite{lowe_distinctive_2004}, which is commonly used for feature extraction and matching between optical images, were extended to be suitable for \ac{sar} images. In \cite{dellinger_sar-sift:_2015}, the SAR-SIFT algorithm has been proposed which focuses on the efficient extraction of local descriptors from \ac{sar} images by modifying the SIFT algorithm to take into account the statistical properties of speckle. However, both of the aforementioned methods are applicable only to \ac{sar} images taken from same-heading orbits with small difference in the respective incidence angles. Specifically for the task of automatic \ac{3D} positioning, in \cite{notarnicola_automatic_2016} the authors have proposed to identify identical scatterers based on detection of local features using the Harris corners. This is followed by constraining the search space by geocoding the local features, using an external digital elevation model and orbit information, and eventually using SIFT for the feature matching. Although the method is promising in terms of detection and positioning of targets even from cross-heading tracks, it only works on isolated \ac{ps}s.\par

In the following we describe in detail the three strategies we apply for detection of identical \ac{ps}s in \ac{sar} images acquired from same- and cross-heading orbits. The methods do not tackle the detection problem directly within the \ac{sar} images but instead rely on external geospatial data and on limiting the search space on geo-referenced positions of the \ac{ps}s. 

\subsection{Multitrack \ac{psi} point cloud fusion}
\label{ssec:fusion}

In \cite{gernhardt_geometrical_2012}, a method for geometrical fusion of multitrack \ac{psi} point clouds stemming from \ac{psi} has been proposed. The fusion algorithm, which is based on the geocoded \ac{psi} point cloud solutions of each geometry as well as information on heading and looking angle of the satellites, consists of three major parts, namely: 1) generation of initial point correspondences, 2) restricted least squares adjustment to minimize the distance between assumed identical points visible from different viewing geometries, and 3) adding a range-dependent shift to all \ac{ps}s using the result of the previous step for the final registration. A summary of the method is described in the following. For a detailed description of the algorithm the reader is referred to \cite{gernhardt_geometrical_2012,gernhardt_high_2012}.\par

Since we are interested in the detection of large number of point correspondences, only the first part of the algorithm is relevant. This coarse registration is performed based on cross-correlation of a subset of geocoded \ac{ps} point clouds from different geometries, after projection on a regular grid, in the xy-, xz- and yz-planes. The subset is chosen based on precision of height update estimates available for each \ac{ps} after carrying out \ac{psi} \cite{gernhardt_high_2012,kampes_radar_2006}. The resulting horizontal and vertical shifts from the mentioned cross-correlation procedure are applied to the \ac{ps}s of one point cloud to align them with the \ac{ps}s of the other point cloud. The coarse shifts are further refined prior to the selection of corresponding \ac{ps} pairs. The refinement is carried out inside a small neighborhood around each \ac{ps} which includes the \ac{ps}s from the other point cloud and tends to accomplish it by performing a statistical search to find the best fit between both \ac{3D} point clouds \cite{gernhardt_high_2012}. The refined shift is applied to the \ac{ps} point cloud of one acquisition geometry and a one-by-one \ac{ps} correspondence is detected in the other point cloud. At the final step, the \ac{3D} coordinates of the geocoded \ac{ps}s have to be projected on the \ac{sar} images of each orbit track, a process called radar-coding. Since the matching of the \ac{psi} results is performed on coordinates in the \ac{utm} map projection, the coordinates are first converted to the Cartesian geocentric system as $(X,Y,Z)_{i}$ for the ith \ac{ps}. Subsequently, the range-Doppler equations described in (\ref{eq:rg_eq}) and (\ref{eq:az_eq}) are inverted to obtain the azimuth and range timing coordinates $(t_{az},\tau_{rg})_{i}$ which can be easily expressed in pixels in the radar coordinate system $(L,P)_{i}$  by knowledge of the range sampling frequency, pulse repetition frequency, first sampled azimuth time and the first sampled range time for each acquisition. The latter information is stated in the product annotation files accompanied by the \ac{tsx} image products \cite{fritz_terrasar-x_2008}.\par

For the same-heading tracks, this method typically generates 200 to 2000 of point correspondences per $\text{km}^{2}$ depending on how densely constructed is a city which directly affects the total number of \ac{ps}s in each point cloud.\par


\subsection{Template matching on optical data}
\label{ssec:template_matching}

Given the availability of suitable remotely sensed optical data, one can detect candidate objects from optical images which are probable to be observed in \ac{sar} images from different viewing geometries. In urban areas, scatterers which are good candidates to be visible from both same-heading and cross-heading tracks usually originate from lamp poles or other cylindrical objects that are vertically oriented towards the sensor. Therefore, the basic idea when using optical data for the aid of \ac{gcp} identification is to detect lamp poles and match the detected objects to the corresponding bright points in \ac{sar} images.\par

The method identifies lamp poles based on their distinctive shadows in optical images using a template matching scheme \cite{montazeri_sar_2016}. Prior to extracting the template, common pre-processing steps such as noise filtering and histogram equalization are carried out on the optical image. Additionally, in order to make the shadows of lamp poles more prominent, a simple sharpening procedure is carried out as follows:

\begin{equation}\label{eq:sharpening}
\mathbf{I} = \mathbf{I}_{o} + a\mathbf{I}_{m},
\end{equation}

where $\mathbf{I}$ is the sharpened image, $\mathbf{I}_{o}$ is the pre-processed original image, $a$ is the scalar sharpening factor and $\mathbf{I}_{m}$ is the un-sharp mask. $\mathbf{I}_{m}$ is calculated as the difference between $\mathbf{I}_{o}$ and its blurred version. Higher values of factor $a$, means higher level of sharpening. The process expressed in (\ref{eq:sharpening}) is called high-boost filtering \cite{gonzalez_digital_2002}.\par

After the sharpening, the template is extracted based on the shadow of an arbitrary lamp pole visible in the optical image. The template is then correlated with the reference image to calculate the following similarity measure for each pixel $(u,v)$ in the reference image \cite{briechle_template_2001}:

\begin{equation}\label{eq:NCC}
\rho (u,v) = \frac{ \Sigma_{x,y} \big[ \mathbf{I}(x,y) - \bar{\mathbf{I}}_{u,v}\big] \big[ \mathbf{T}(x-u, y-v) - \bar{\mathbf{T}} \big] }{\sqrt{\Sigma_{x,y} \big[\mathbf{I}(x,y) - \bar{\mathbf{I}}_{u,v}\big]^{2}~\Sigma_{x,y} \big[ \mathbf{T}(x-u, y-v) - \bar{\mathbf{T}} \big]^2}},
\end{equation}

where $\mathbf{I}(x,y)$ and $\mathbf{T}(x,y)$ denote pixel values of the reference and the template image at $(x,y)$, respectively, and $\Sigma_{x,y}$ stands for $\Sigma_{x=1}^{N_{1}} \Sigma_{y=1}^{N_{2}} $ with $N_{1} \times N_{2}$ being the size of the template. Furthermore,  $\bar{\mathbf{I}}_{u,v}$ and $\bar{\mathbf{T}}$ denote the mean intensity values of the original image and the template, respectively. Equation (\ref{eq:NCC}) allows for calculation of the \ac{ncc} value $\rho (u,v)$ which leads to the detection of the template location in the reference image after proper thresholding. It is important to note that due to the normalization carried out in the denominator of (\ref{eq:NCC}), $\rho (u,v)$ is independent of changes in brightness or contrast of the image and therefore improves the result of template matching.\par

After detection of pixels which belong to the shadows of lamp poles, the result is geo-referenced in the \ac{utm} coordinate system. Since for each lamp pole in the optical image more than one pixel exists which represent the object, a subsequent clustering is performed. The clustering is carried out non-parametrically using the mean shift concept \cite{comaniciu_mean_2002}:

\begin{equation}\label{eq:mean_shift}
\mathbf{M}(\mathbf{p}_{i}) = \frac{ \Sigma_{j=1}^{n} ~ \mathbf{p}_{j} ~ g \Big( \lVert \frac{\mathbf{p}_{i} - \mathbf{p}_{j}}{h} \rVert^{2} \Big) }{\Sigma_{j=1}^{n} ~ g \Big( \lVert \frac{\mathbf{p}_{i} - \mathbf{p}_{j}}{h} \rVert^{2} \Big)} - \mathbf{p}_{i},
\end{equation}

where $\mathbf{p}_{i}$ denotes a \ac{3D} point for which the shift vector $\mathbf{M}(\mathbf{p}_{i})$ is calculated. $\mathbf{p}_{j}$ represents the points in a neighborhood of $\mathbf{p}_{i}$, $g$ is a kernel function with the bandwidth $h$ and $\lVert \cdot \rVert$ is the Euclidean distance operator. The main idea of the algorithm is to shift each point in a small neighborhood towards its weighted mean value and thus representing each cluster by its centroid \cite{comaniciu_mean_2002}. The process is carried out iteratively until the length of $\mathbf{M}(\mathbf{p}_{i})$ becomes equal or close to zero. For our application, since in any case there will be a mismatch between the detected points on optical data and the corresponding bright points in the \ac{sar} image, utilizing a flat kernel in equation (\ref{eq:mean_shift}) suffices. This means the algorithm is simplified by calculating the sample mean in a specified radius of $\mathbf{p}_{i}$ and shifting the desired point towards the estimated center.\par

In the next step, the clustered points with \ac{utm} coordinates should be radar-coded to all the available \ac{sar} images. As it was mentioned earlier, the positions of the detected lamp poles on the optical data and the bright \ac{ps}s in the SAR image will most probably not coincide after radar-coding. This can be explained by height uncertainties of the geo-referenced optical data and the fact that the data may not be perfectly orthorectified. Therefore in the final step of the algorithm, the detected lamp poles are registered on the corresponding bright dots in the \ac{sar} image by employing the iterative closest point (ICP) algorithm \cite{besl_method_1992}. To this end, binary masks are generated based on thresholding on the bright points on the \ac{sar} image and keeping only the detected lamp poles from the optical data. The ICP algorithm then finds for each individual point its closest point in the corresponding point set. It iteratively estimates the transformation parameters (translation and rotation) to minimize the mean squared error between the two point sets and finally registers one point cloud onto the other point cloud with the refined transformation parameters.\par

It is noteworthy that the positioning accuracy of the utilized optical image does not necessarily have to be high. A horizontal positioning accuracy in the order of couple of meters and an approximate knowledge of height based on freely available sources usually suffice for the procedure described in this subsection. If a mismatch occurs due to the low positioning accuracy, this will be compensated by the final step of the algorithm with applying the ICP. However, the spatial resolution of the optical data should be strictly high, 10 cm or better, in order to be able to accurately detect the shadows of the lamp poles.


\subsection{Vector road network data}
\label{ssec:road_network}

In urban areas, the cylindrical objects of our interest (lamp poles, road signs, traffic lights, etc.) are typically located along the roads. Therefore, with the availability of geospatial road data, either obtained from OpenStreetMap or country-specific geoportals, and the projection of such maps on \ac{sar} images, one can search for bright points in the neighborhood of the road data points.\par

The method is applied to co-registered stacks of \ac{sar} images. If \ac{sar} stacks from multiple viewing geometries are available, first the road data, which is usually delivered in the \ac{utm} coordinate system, is radar-coded based on the master orbit information of each stack. It is important to note that if the road data do not have any information about the ellipsoidal height, then for the radar-coding a constant height value based on prior knowledge can be chosen for all the road data points. Furthermore, the data with horizontal positioning accuracy in the order of couple of meters will suffice for the PS matching procedure as the PS correspondences are detected on a neighborhood-analysis basis, as is depicted in the following, which does not require the exact position of the road data point.\par

After radar-coding, a circular neighborhood is considered around each road data point. The radius of the circle depends on the typical width of streets and highways. Subsequently, for each pixel within the neighborhood the \ac{adi} is evaluated \cite{ferretti_permanent_2001}:

\begin{equation}\label{eq:ADI}
D_{a} \approx \frac{\sigma_{a}}{\bar{a}},
\end{equation}

where $\sigma_{a}$ and $\bar{a}$ are the temporal standard deviation and the temporal mean of calibrated amplitude values of the pixel, respectively, and $D_{a}$ approximates the phase dispersion. The pixel with the lowest value of $D_{a}$, i.e. the one with the highest phase stability is chosen as potential \ac{ps} candidate. This process is carried out for all of the available road data points. Since at this point, it is possible that many false pixels with relatively low $D_{a}$ values in the neighborhoods are categorized as potential \ac{gcp} candidates, a further thresholding on $D_{a}$ is performed in \ac{sar} images from all available viewing geometries. This operation, in addition to constraining the approximate elevation of the \ac{ps} candidates to be close to the ground, causes a dramatic decrease in the total number of detected candidates but improves the accuracy of the detection.\par

Finally, the presumable identical \ac{ps}s in all available geometries are geocoded using the respective master orbit information. In the geocoded results the \ac{ps}s which are close enough, in terms of coordinate differences, are selected as the final \ac{gcp} candidates.\par


\section{Automatic \ac{gcp} Generation: The Processing Chain}
\label{sec:chain}

\begin{figure*}[!t]
        \includegraphics[width=0.95\textwidth]{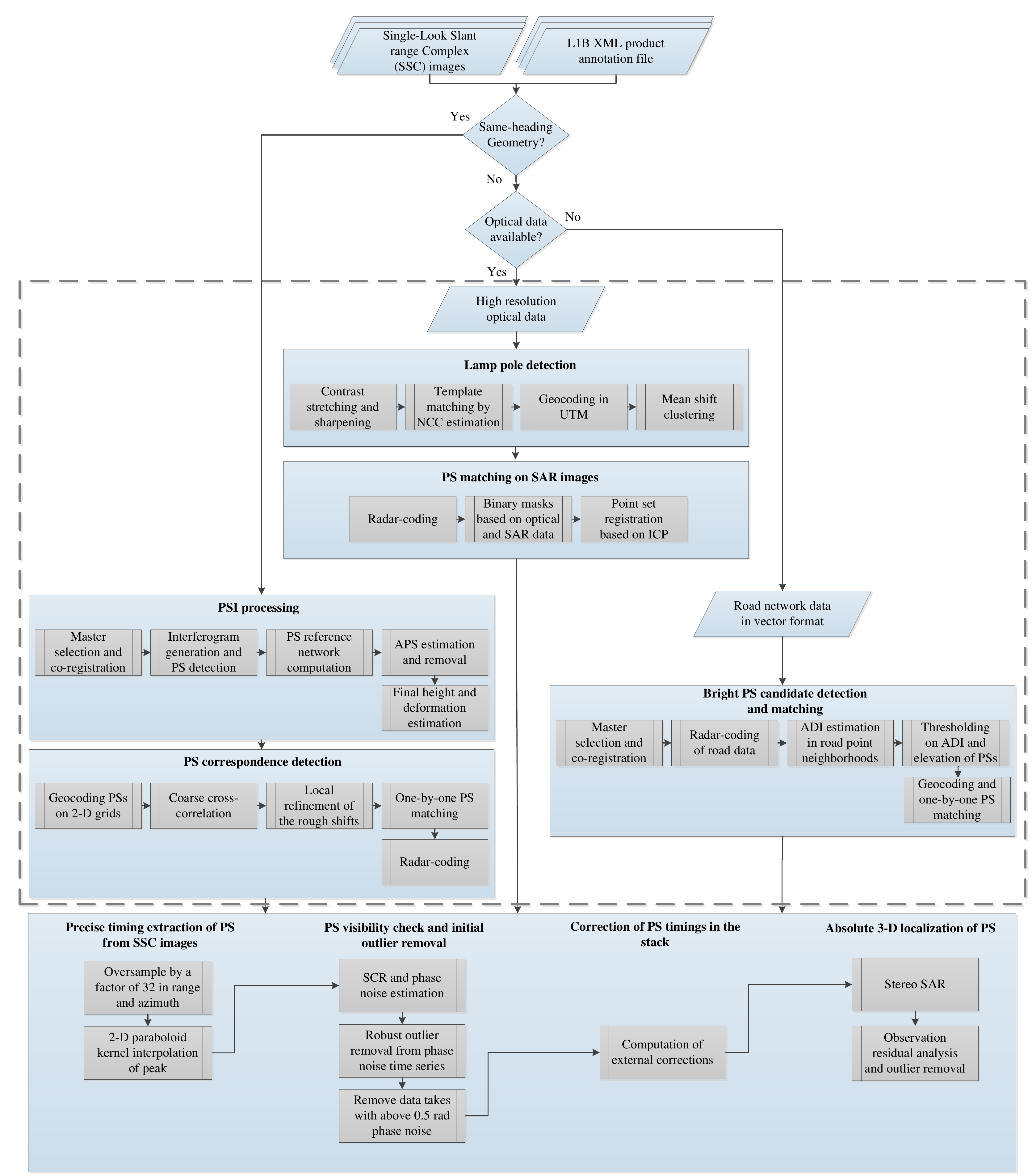}
        \caption{The flowchart of the automatic \ac{gcp} generation algorithm. The input data is the Single-Look Slant range Complex (SLC) SAR images and their accompanying L1B product annotation files. The identical \ac{ps} detection part is drawn inside the gray dashed rectangle. In case of \ac{ps} detection from cross-heading viewing geometries, the algorithm receives as input the optical image or road network data of the scene upon availability. The processes are depicted with big blue rectangles while the sub-processes are shown with small gray rectangles. The name of each process step is written in bold letters. The last blue rectangle includes the processes and sub-processes that all the detected candidates should go through independent of the detection methodology.}\label{fig:gcp_flowchart}
\end{figure*}

The processing chain for automatic detection and positioning of \ac{gcp}s includes a set of procedures which starts from single-look slant range complex \ac{sar} images and their corresponding product annotation files to absolute \ac{3D} coordinates of the chosen \ac{gcp}s. The flowchart of the algorithm is shown in Fig. \ref{fig:gcp_flowchart}. It consists of the following major processes which have to be carried out in the stated order:

\begin{enumerate}
\item identification of identical scatterers visible in multi-aspect \ac{sar} images.
\item precise extraction of scatterers' azimuth and range positions from \ac{sar} images at sub-pixel level.
\item scatterer visibility check and initial removal of outliers from time series of phase noise.
\item correction of radar timings for all the perturbing signals.
\item estimation of the \ac{3D} absolute coordinates of the scatterers.
\end{enumerate}

This section discusses each step. It is important to note that since some of the steps are well-established and well-documented techniques in the \ac{sar} community, all the details will not be repeated here. Instead, for these processes, the relevant references are provided.

\subsection{Identification of identical \ac{ps}}
\label{ssec:identification_PS}

This part has been already covered in Section \ref{sec:detection} and is shown as the processes inside the gray dashed rectangle in the flowchart of Fig. \ref{fig:gcp_flowchart}. For detection of \ac{ps} candidates from same-heading orbits, the \ac{psi} processing is carried out following the guidelines from \cite{kampes_radar_2006,ferretti_permanent_2001,adam_development_2003,adam_wide_2013} and the identical \ac{ps} matching is done using the \ac{psi} multitrack fusion algorithm described in \cite{gernhardt_geometrical_2012,gernhardt_high_2012}. In the case of localization from cross-heading tracks, the \ac{ps} candidate selection is carried out using the methods based on the optical data or on the road network data. Regardless of the detection method, the output of this step is the approximate radar coordinates of identical \ac{ps}s in terms of lines and samples in all non-coregistered images of different orbit tracks.

\subsection{Precise extraction of \ac{ps} timings}
\label{ssec:PTA}

The rough radar coordinates of the \ac{ps}s from the previous step should be refined to sub-pixel level in order to extract the timings precisely. To this end, a process called \ac{pta} is carried out \cite{cumming_digital_2005}. In each image of the scene, a 32 $\times$ 32 window centered on the \ac{ps} is extracted. In both range and azimuth direction, an oversampling by a factor of 32 is performed and the integer peak position of the \ac{ps} response is measured. Subsequently in a 3 $\times$ 3 window centered on the peak position, a paraboloid interpolation is performed to refine the values around the maximum. This method allows the retrieval of the peak position of the \ac{ps} with a sensitivity better than $\frac{1}{1000}$ of a pixel in each dimension \cite{balss_high_2012}. These values are then converted to radar timings based on the product annotation files \cite{fritz_terrasar-x_2008}.\par

Based on the result of the \ac{pta} for each \ac{ps}, the refined peak power $P_{peak}$ and the clutter power $P_{clutter}$ can be computed. These values, if expressed in $dB$, are related to the \ac{scr} of the target as:

\begin{equation}\label{eq:SCR}
\text{SCR} = 10^{\big( \frac{P_{peak} - P_{clutter}}{10} \big) }
\end{equation}

which is expressed as a digital number.

%
%
%
%

\subsection{\ac{ps} visibility check and initial outlier removal}
\label{ssec:outlier_rem}

The \ac{scr} of potential \ac{ps} candidates should be high enough in the stack of \ac{sar} images so that the \ac{ps} can be localized with high precision. Therefore, we analyze the time series of phase noise values $\bm{\sigma_{\phi}}$ of the \ac{ps}s to exclude potential outliers and check if the scatterer is visible in one data take or not. The $\sigma_{\phi}$ of a  \ac{ps} in acquisition $i$ is related to the \ac{scr} of the target as \cite{adam_development_2004}:

\begin{equation}\label{eq:phase_nouise}
\sigma_{\phi_{i}} \approx \frac{1}{\sqrt{2~\text{SCR}_{i}}} ~
\end{equation}

which is expressed in radians. The values of $\bm{\sigma_{\phi}}$, for a specific \ac{ps} in all data takes, are non-negative and follow a right-skewed distribution. Removal of outliers based on statistical measures such as mean or median is not recommended since many regular values can be categorized as outliers. Therefore, we use a method called \textit{adjusted boxplot} which allows for robust elimination of outliers in univariate skewed distributions \cite{hubert_adjusted_2008}.\par

The main idea of the adjusted boxplot is to modify the original boxplot method, described in \cite{tukey_exploratory_1977}, to include information about the skewness of the data. Therefore, instead of classifying an observation as outlier if it lies outside of the interval defined by the boxplot method \cite{tukey_exploratory_1977}:

\begin{equation}\label{eq:bp}
\big[ Q_1-1.5 ~ IQR ; ~ Q_3+1.5~ IQR \big],
\end{equation}

the adjusted boxplot method declares an observation as outlier if its value exceeds the following interval \cite{hubert_adjusted_2008}:

\begin{equation}\label{eq:bp_ad}
\big[Q_{1} - 1.5 ~ e^{(-4MC)} ~ IQR; ~ Q_{3} + 1.5 ~ e^{(3MC)} ~ IQR \big].
\end{equation}

In (\ref{eq:bp}), $Q_1$ and $Q_3$ are the first and the third quartiles of the data, respectively and $IQR=Q_{3}-Q_{1}$ denotes the interquartile range. In (\ref{eq:bp_ad}), $MC$ is the medcouple, a robust measure of the skewness of a univariate sample which for right-skewed distributions is always non-negative \cite{brys_robust_2004}. The exponential functions in (\ref{eq:bp_ad}) which depend on the $MC$ as well as the included coefficients are chosen experimentally based on some well-known skewed distributions. For more details on the theory and implementation of the adjusted boxplot method, the reader is referred to \cite{hubert_adjusted_2008}.\par

After the automatic identification and exclusion of $\sigma_{\phi}$ values which do not lie within the interval of (\ref{eq:bp_ad}), the remaining time series is analyzed to remove the data takes in which the specific \ac{ps} is not visible. This is done by removing all $\sigma_{\phi}$ values which are above 0.5 radians ($\approx 30^{\circ}$) as is stated in \cite{kampes_radar_2006}.

\subsection{Correction of \ac{ps} timings in a stack}
\label{ssec:corrections}

The correction of the extracted \ac{ps} timings is performed using the imaging geodesy technique \cite{eineder_imaging_2011} which was briefly introduced in Subsection \ref{ssec:backgr}. It is worth mentioning that the tropospheric and ionospheric effects are corrected based on global numerical weather models and global ionospheric maps, respectively, if local \ac{gnss} receivers are not available in the vicinity of the investigated area. Along with these corrections comes the corresponding geometrical calibration of the \ac{sar} sensor in range and azimuth which ensures centimeter localization accuracy. The calibration is based on corner reflectors with known reference coordinates \cite{eineder_imaging_2011}. The output of this part is the absolute \ac{2D} radar timing coordinates.

\subsection{Absolute \ac{3D} localization of \ac{ps}}
\label{ssec:stereo_3D}

At the final step in the processing chain, the corrected range and azimuth timings from the entire multi-aspect set of \ac{sar} images are combined to retrieve the absolute \ac{3D} position of the \ac{ps} with the stereo \ac{sar} method described in Subsection \ref{ssec:backgr}. Apart from the \ac{3D} position of the target, stereo \ac{sar} reports on the standard deviation of each coordinate component ($S_X$, $S_Y$, $S_Z$) as the by-product of the least squares adjustment. Furthermore, observation quality of each \ac{ps} i.e. the azimuth and range standard deviations ($S_{az}$, $S_{rg}$), retrieved from applying VCE to residuals, are delivered. It is important to note that the VCE is carried out individually for each geometry which allows to judge the consistency of the observed geometries with respect to the underlying assumption that the intersection occurs at a common \ac{ps}.\par

The residuals of the adjusted range and azimuth observations are the basis for the elimination of outliers after stereo \ac{sar} processing. Therefore, the processing is carried out repeatedly, where first the initial \ac{3D} coordinates are estimated using the provided input timings. The range and azimuth residuals are analyzed to exclude observations with residual values larger than 0.6 m in range or larger than 1.1 m in azimuth. The upper bounds correspond to the nominal spatial resolution of \ac{tsx} high resolution spotlight products used in this study. Then stereo \ac{sar} is performed again with the cleaned observations. This time, observations which show residuals larger than two times the $S_{az}$ and the $S_{rg}$ are removed. Additionally, to remove the \ac{ps}s which are not considered ideal for stereo \ac{sar} due to wrong correspondence matching caused by several scatterers being too close, a third step of data cleaning is performed. \ac{ps}s having an $S_{az}$ higher than 20 cm in any of their azimuth geometries are removed based on the assumption that the discrepancy should not exceed the typical size of the \ac{ps} object, for instance, a lamp pole.\par

The estimated variance-covariance matrix of the \ac{3D} position of the target is further used for error analysis. The matrix gives important information about the stability of the coordinates results and is affected by the factors stated in Subsection \ref{ssec:backgr}.

\section{Experimental Results}
\label{sec:results}

In this section, the work-flow described in Section \ref{sec:chain} is applied on real data to produce remotely sensed \ac{sar}-based \ac{gcp}s. In Subsection \ref{ssec:berlin}, the results are reported for a small test site in Berlin where the detection of \ac{gcp} candidates are carried out using cross-heading geometries. In Subsection \ref{ssec:oulu}, the processing results are shown for the entire city of Oulu, Finland, where the detection and positioning are performed on \ac{ps} candidates detected from both same- and cross-heading orbit tracks using the methods described in Subsections \ref{ssec:fusion} and \ref{ssec:road_network}, respectively.

\subsection{Berlin}
\label{ssec:berlin}

The first test site includes a small area close to the Berlin central railway station. The \ac{sar} data set, 214 images in total, consists of two stacks of \ac{tsx} very high resolution spotlight images acquired with a range bandwidth of 300 MHz. The images cover a period of approximately six years from April 2010 to March 2015. In terms of viewing geometry, one stack was acquired from a descending orbit with images recorded at 05:20 coordinated universal time (UTC), and one stack was acquired from an ascending track with images recorded at 16:50 UTC. The acquisition parameters of each stack are summarized in Tab. \ref{tab:acq_berlin}.

\begin{table}[tbh!]
\centering
\ra{1.6}
\caption{\textsc{Acquisition parameters of stacks of \ac{sar} images in Berlin}}
  \begin{tabulary}{0.48\textwidth}{|C|CCCC|}
    \hline
    \textbf{Beam Nr.} & \textbf{Incidence angle (degree)} & \textbf{Heading angle (degree)} & \textbf{Track type} & \textbf{Nr. of images} \\
    \hline
    57     & 41.9                   & 350.3                & Asc  & 107          \\
    \hline
    42     & 36.1                   & 190.6                & Dsc  & 107           \\ \cline{1-5}
  \end{tabulary}
  \label{tab:acq_berlin}
  \end{table}

\begin{figure*}[tbh!]
\centering
        \includegraphics[width=0.9\textwidth]{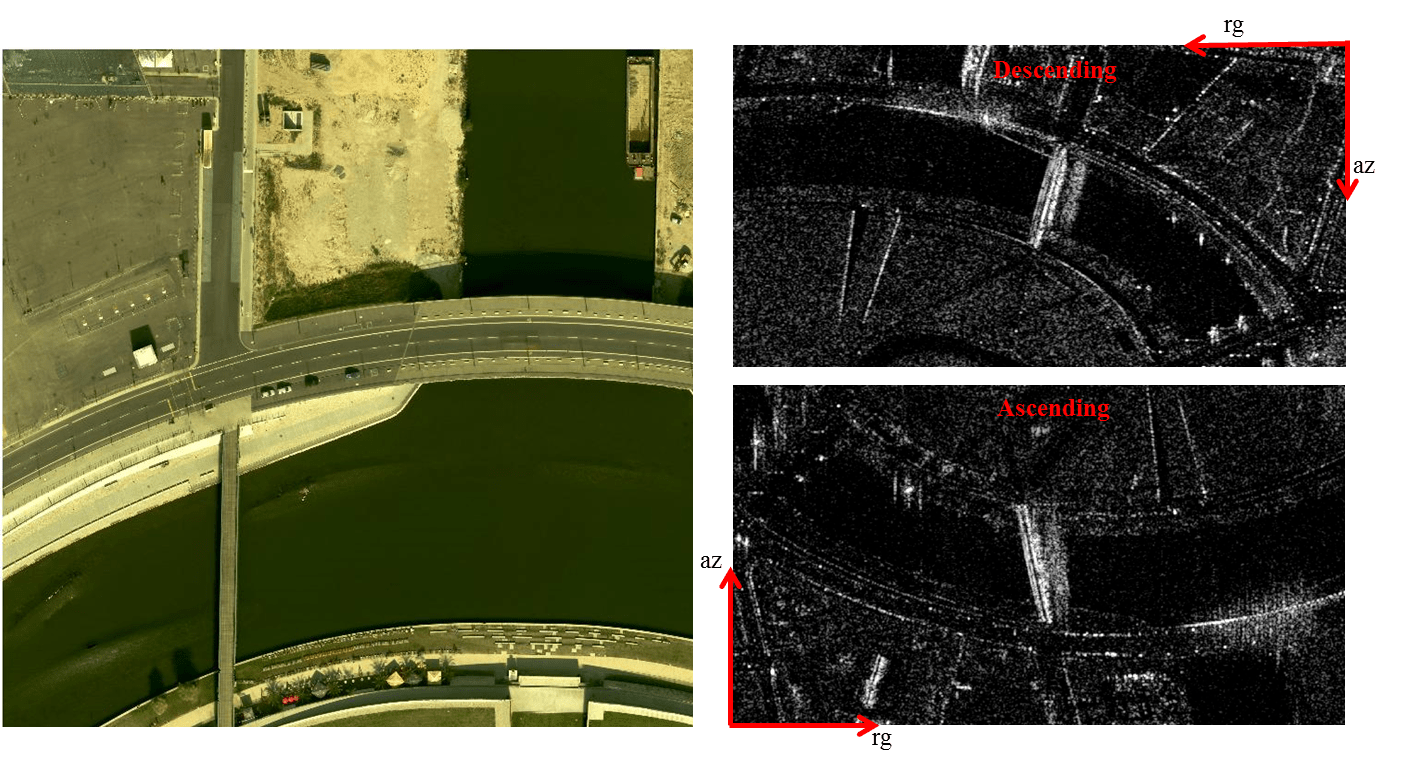}
        \caption{The optical image (left) and the \ac{sar} amplitude images (right) of the test area in Berlin. The contrast of the optical image has been adjusted to illustrate the shadows of the lamp poles prominently.}\label{fig:opt_sar_Berlin}
\end{figure*}

For the selected test site, an aerial optical image with ground spacing of 7 cm is also available. The optical image is orthorectified and was used in a stereo matching process to produce a digital surface model with decimeter accuracy \cite{hirschmuller_stereo_2008}. The optical image of the test site and the corresponding \ac{sar} images are shown in Fig. \ref{fig:opt_sar_Berlin}.

\begin{figure*}[!tbh] 
\begin{subfigure}{0.3\textwidth}
\includegraphics[width=\linewidth]{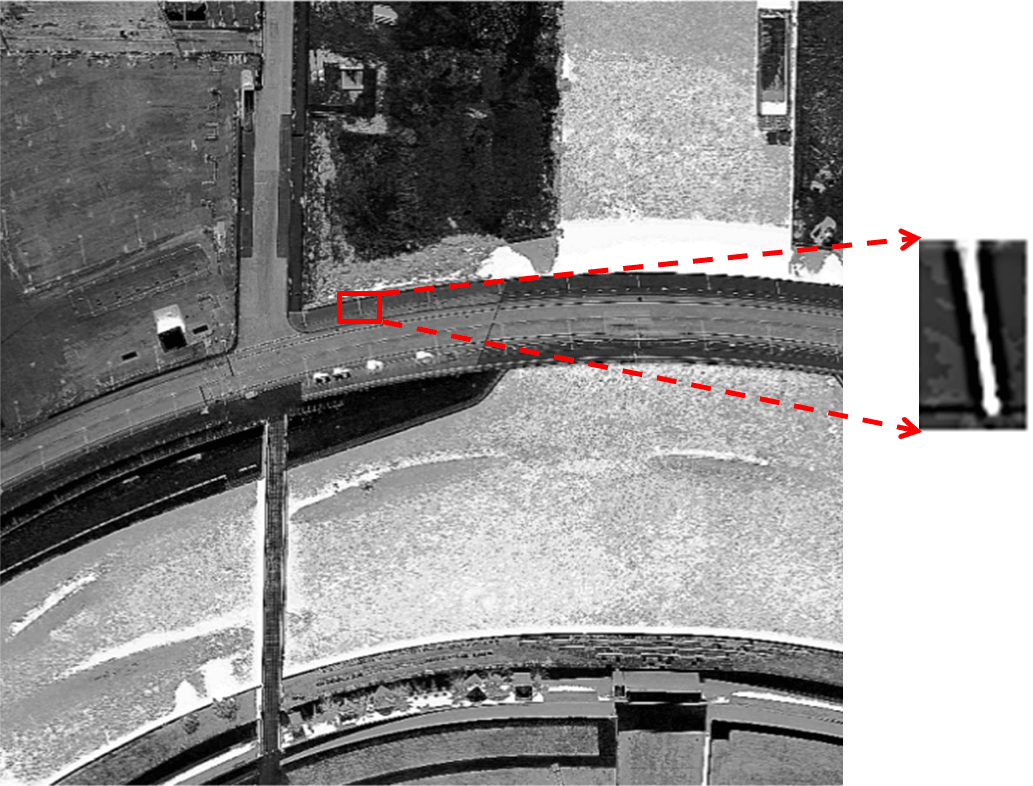}
\caption{} \label{fig:sharp_temp}
\end{subfigure}\hspace*{\fill}
\begin{subfigure}{0.3\textwidth}
\includegraphics[width=\linewidth]{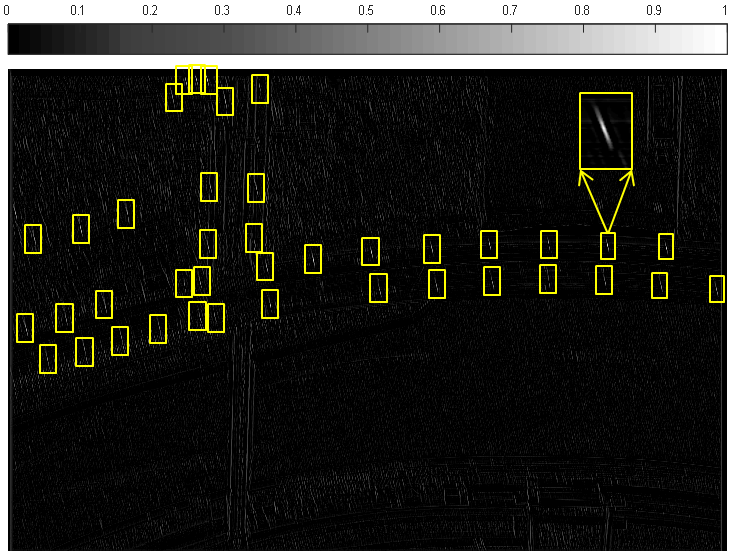}
\caption{} \label{fig:NCC_marked}
\end{subfigure}\hspace*{\fill}
\begin{subfigure}{0.3\textwidth}
\includegraphics[width=\linewidth]{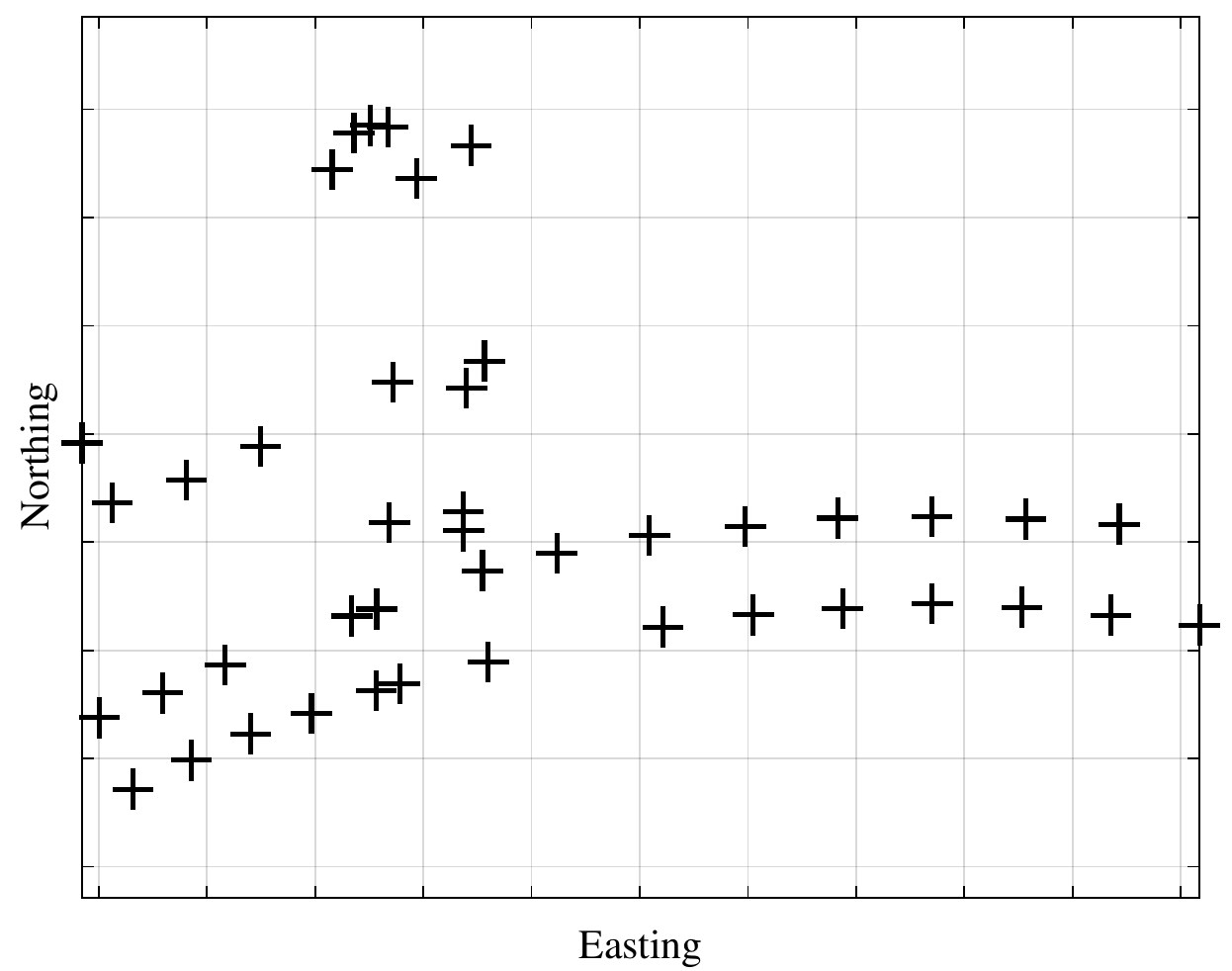}
\caption{} \label{fig:utm_lps}
\end{subfigure}

\medskip
\begin{subfigure}{0.5\textwidth}
\includegraphics[width=\linewidth]{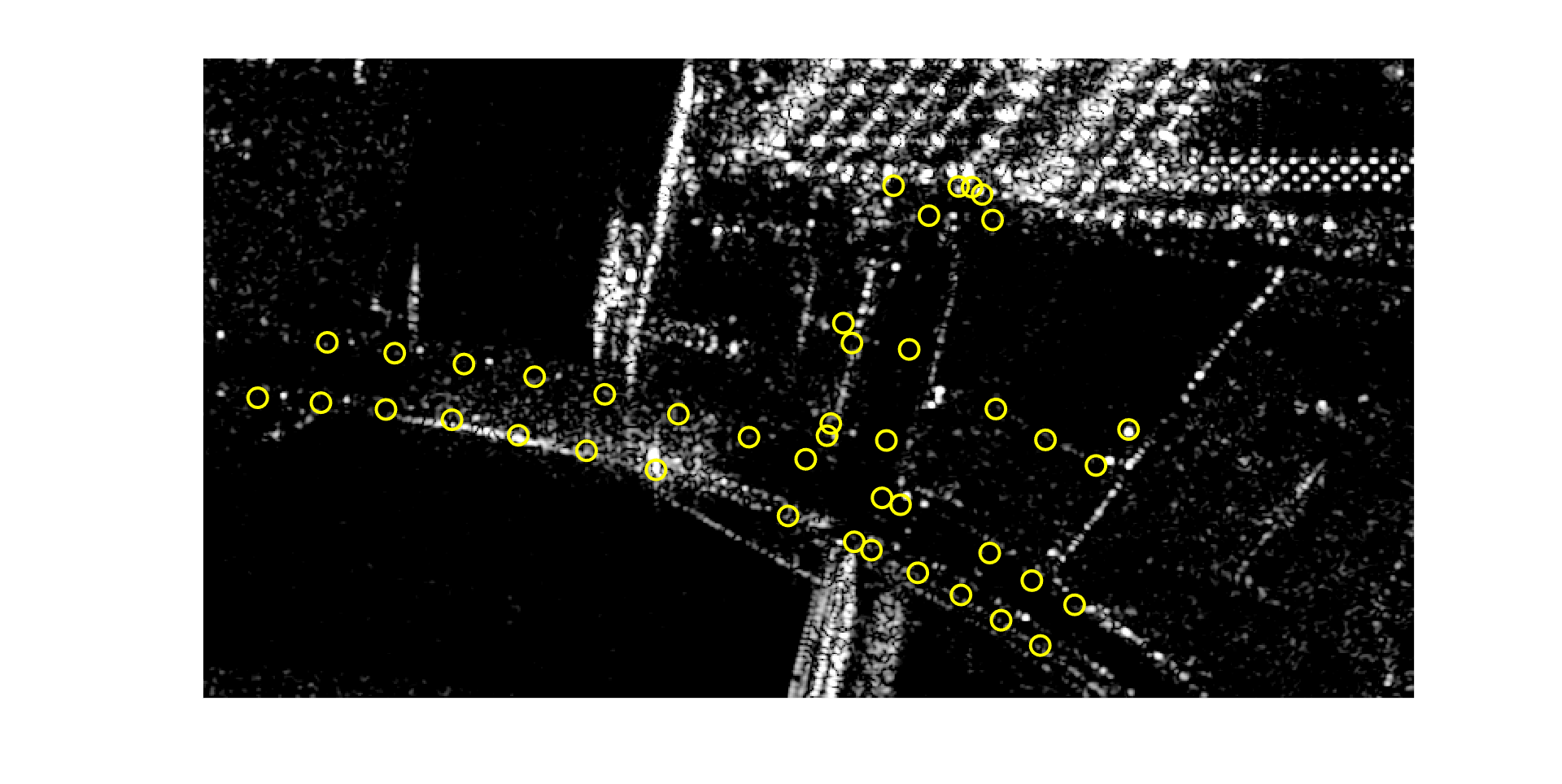}
\caption{} \label{fig:dsc_bef_ICP}
\end{subfigure}\hspace*{\fill}
\begin{subfigure}{0.5\textwidth}
\includegraphics[width=\linewidth]{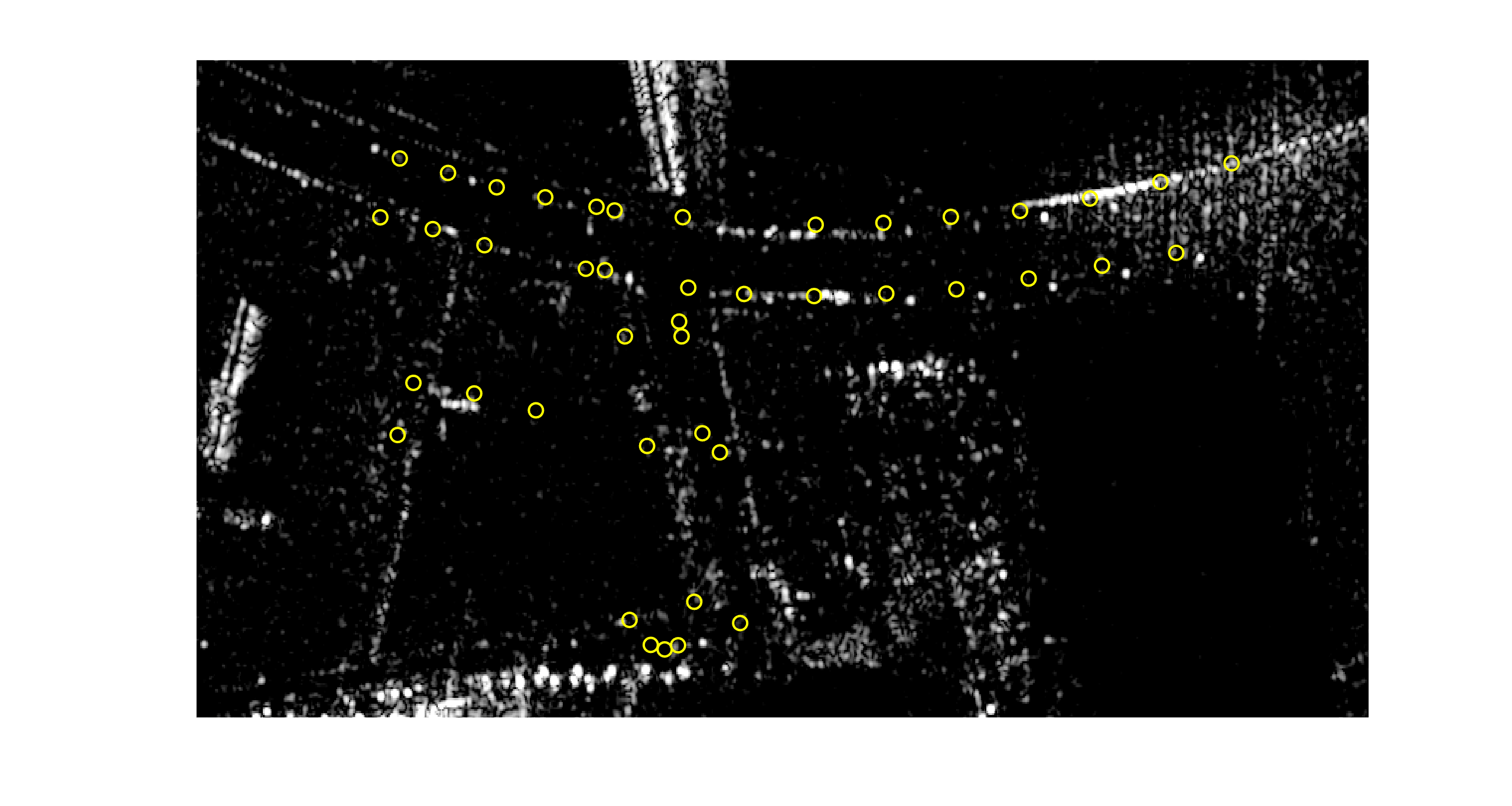}
\caption{} \label{fig:asc_bef_ICP}
\end{subfigure}

\begin{subfigure}{0.5\textwidth}
\includegraphics[width=\linewidth]{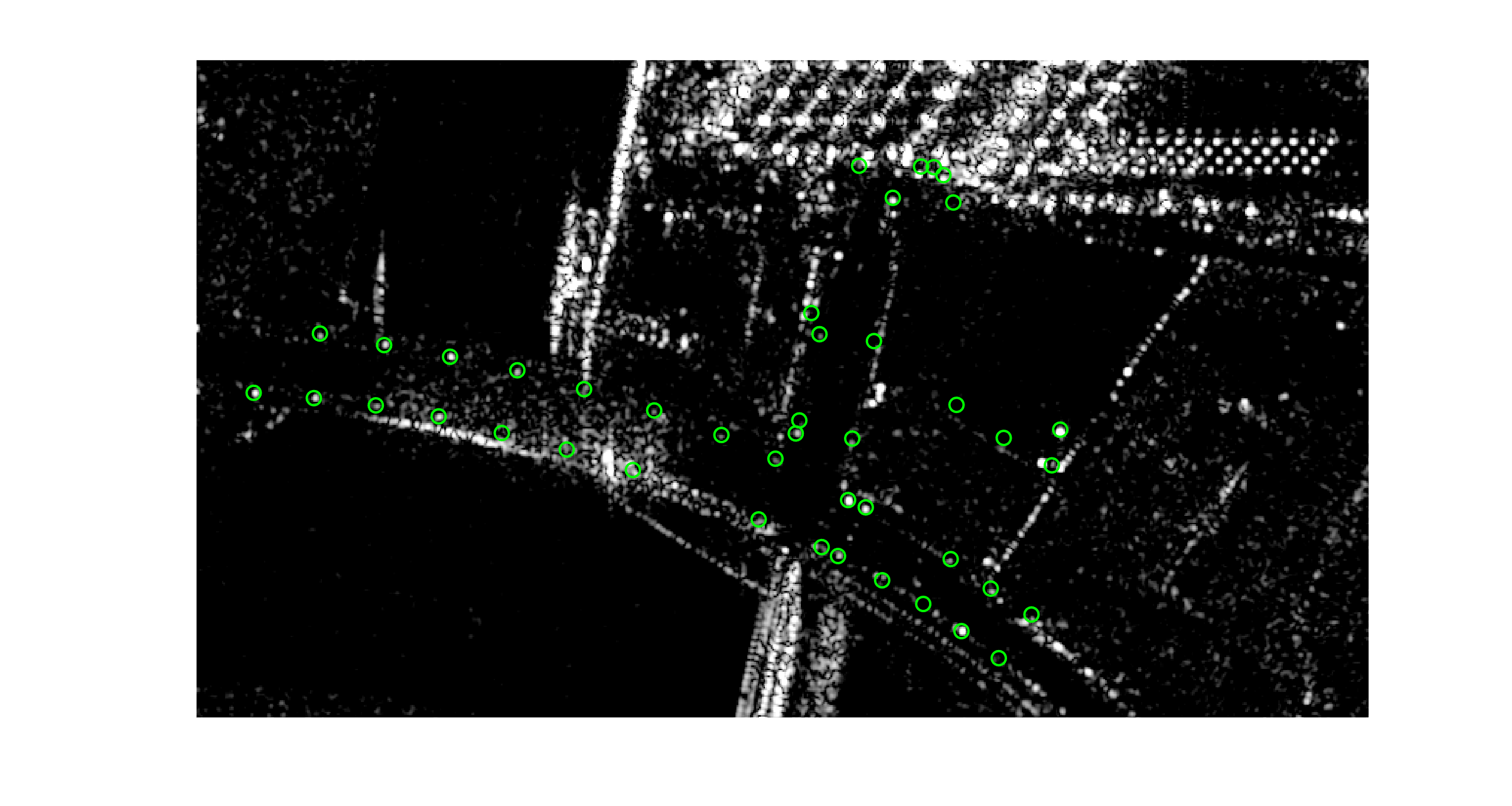}
\caption{} \label{fig:dsc_aft_ICP}
\end{subfigure}\hspace*{\fill}
\begin{subfigure}{0.5\textwidth}
\includegraphics[width=\linewidth]{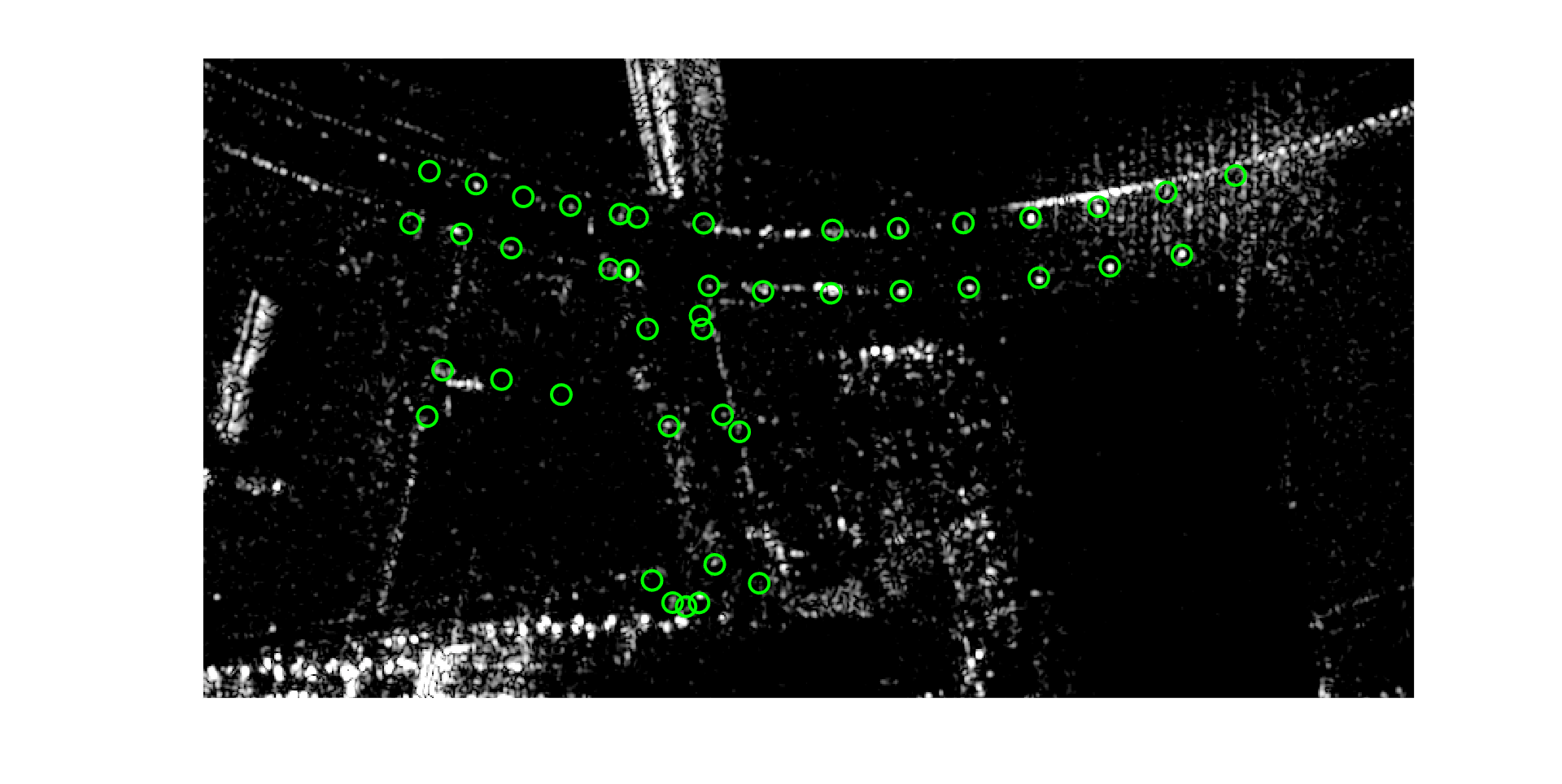}
\caption{} \label{fig:asc_aft_ICP}
\end{subfigure}

\caption{Demonstration of \ac{ps} correspondence detection in Berlin based on high resolution optical data. (a) shows the pre-processed optical image after negative intensity transformation and the extracted template. (b) is the calculated \ac{ncc} map after correlating the extracted lamp pole template with the reference image in which the detected objects are marked by yellow rectangles. (c) shows the 44 detected objects in the \ac{utm} coordinate system after clustering. In (d) and (e), the radar-coded results are depicted by yellow circles which show offsets with respect to the bright points in the \ac{sar} images. (f) and (g) show the results of matching after using the ICP algorithm on the descending and the ascending image, respectively. In the last two subfigures, it can be seen that the detected objects from the optical image (green circles) coincide with the visible bright points in the \ac{sar} images.}\label{fig:Berlin_PS_Matching}
\end{figure*}

The \ac{gcp} candidate selection is carried out based on the optical data which includes the detection of lamp poles and their projection onto cross-heading \ac{sar} images (see Subsection \ref{ssec:template_matching}). The individual steps of this process, applied on the Berlin test site, are shown in Fig. \ref{fig:Berlin_PS_Matching}. After the extraction of the template (see Fig. \ref{fig:sharp_temp}), the \ac{ncc} map is calculated and pixels with values higher than 0.6 are classified as parts of shadows of lamp poles as illustrated in Fig. \ref{fig:NCC_marked}. It is important to note that the exact tuning of the threshold is not necessary as long as the value is chosen low enough. If the threshold is strictly chosen as a high value, although we are selecting the most similar pixels to the template, we may ignore all the other candidates which show less similarity to the template but might have been potential candidates for stereo SAR processing and \ac{3D} localization. Therefore, in our processing chain the default value is set to 0.6, which is slightly higher than the half of the \ac{ncc} range [0,1], to guarantee a certain degree of similarity while selecting a large number of pixels. In the Berlin case study, this leads to the selection of 2030 pixels which are further clustered to represent the 44 detected objects in the \ac{utm} coordinate system (see Fig. \ref{fig:utm_lps}). The objects include the lamp poles along the bridge and at the street perpendicular to the bridge as well as flag poles at the top left of the optical image close to the Berlin railway station. After radar-coding of the results onto the entire ascending and descending \ac{sar} images, the mismatch between the projected points and the actual bright points in the \ac{sar} images, depicted in Fig. \ref{fig:dsc_bef_ICP} and Fig. \ref{fig:asc_bef_ICP}, is resolved using ICP. The detection outcome is marked with green circles in Fig. \ref{fig:dsc_aft_ICP} and Fig. \ref{fig:asc_aft_ICP}. It is seen from the results that not all of the available street lights can be detected using the mentioned strategy as some can be occluded by cars or the object's shadow is not distinctive enough to match the extracted template. However, this is of low importance in our application since in such a small test site, with an area less than two $\text{km}^{2}$, two or three \ac{gcp}s will certainly suffice. Moreover, if the method detects wrong candidates which do not fall in the category of \ac{ps}s, the subsequent \ac{pta} and phase noise analysis will discard these points.\par

The precise radar timings of the \ac{ps}s are extracted using \ac{pta} (see Subsection \ref{ssec:PTA}). Subsequently for each \ac{ps} candidate in each data take, \ac{scr} and $\sigma_{\phi}$ values are evaluated using (\ref{eq:SCR}) and (\ref{eq:phase_nouise}), respectively. After excluding potential outliers with the adjusted boxplot method, the data takes in which the scatterer is not visible are discarded by thresholding on the $\sigma_{\phi}$ values (see Subsection \ref{ssec:outlier_rem}). At this point, one might argue that analyzing the time series of the remaining $\sigma_{\phi}$ values can already give a hint if the scatterer is a suitable candidate for positioning or not. This statement is partially true since the mentioned analysis is only useful to separate the time-coherent scatterers from the non-coherent ones. It is possible that several scatterers, located close to each other, are mapped as one bright point which results in low $\sigma_{\phi}$ value but of course not a suitable candidate for stereo \ac{sar}. These candidates are usually discarded in the stereo \ac{sar} processing due to large $S_{az}$ values which indicate that not the same object has been detected from multiple viewing geometries. Therefore, in our processing chain, the detected candidates are not entirely removed based on their average \ac{scr} and they are all passed to the final stereo \ac{sar} processing.\par

As for the corrections, the ionospheric delay is estimated using global ionospheric maps. The tropospheric delay is estimated using the zenith path delay information of the closest permanent \ac{gnss} station in Potsdam which is situated approximately 35 km away from the test site.

\begin{figure*}
\centering
   \begin{subfigure}[!tbh]{0.9\textwidth}
   \includegraphics[width=1\linewidth]{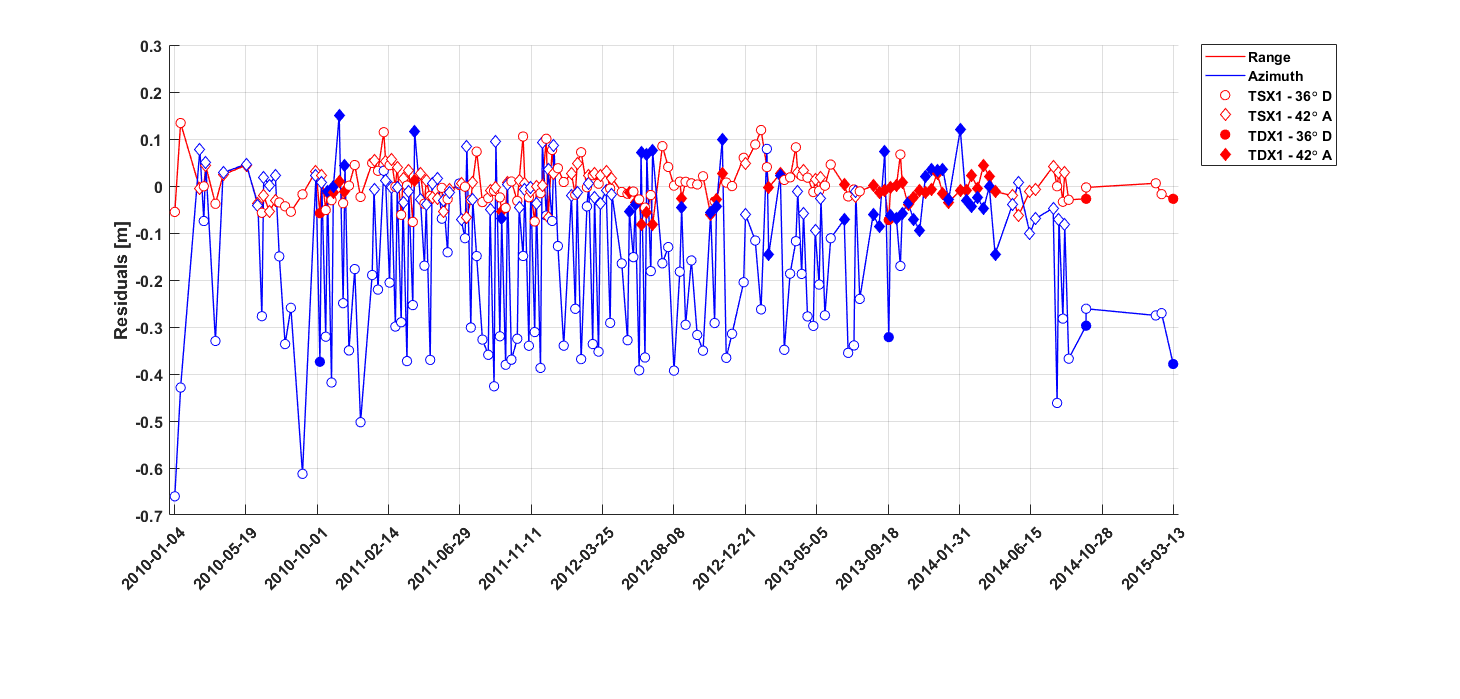}
   \caption{$\text{PS}_{41}$}
   \label{fig:residual_plots1}
\end{subfigure}
\medskip
\begin{subfigure}[!thb]{0.9\textwidth}
   \includegraphics[width=1\linewidth]{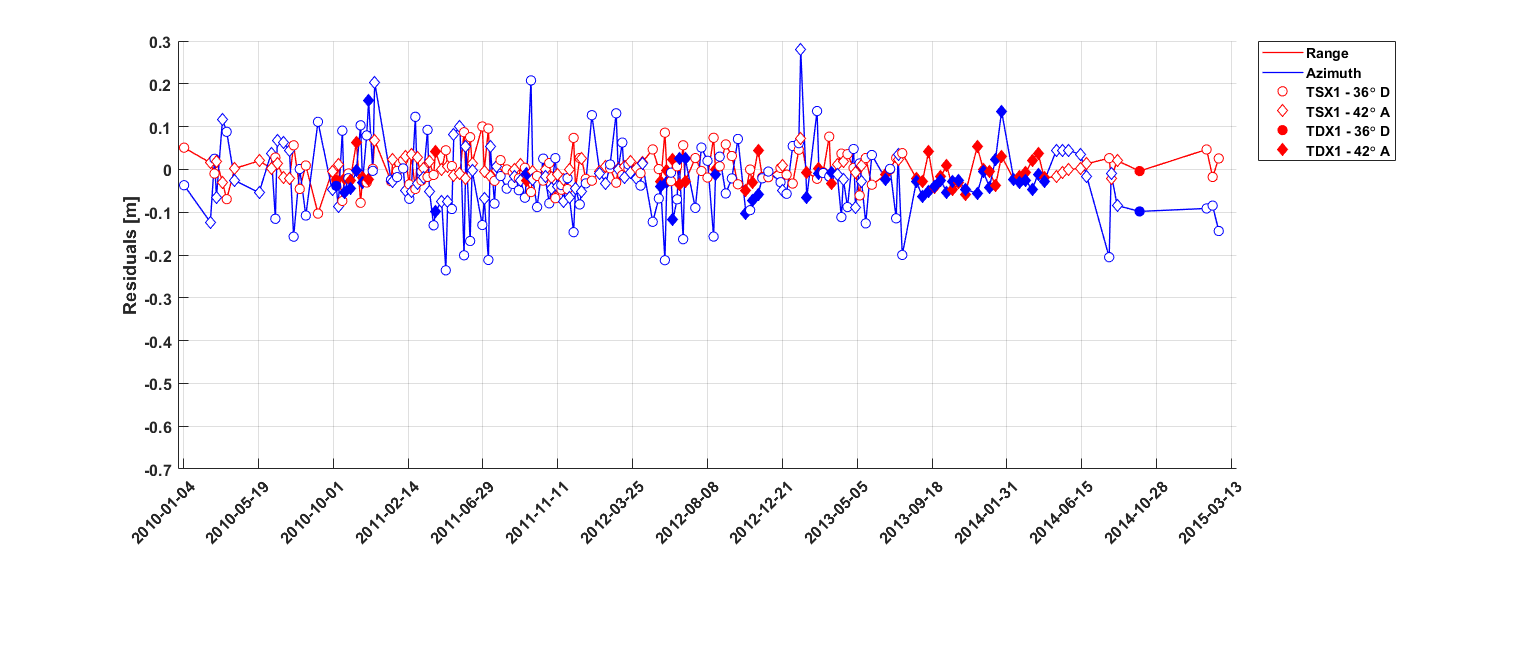}
   \caption{$\text{PS}_{43}$}
   \label{fig:residual_plots2}
\end{subfigure}
\caption{Range and azimuth residuals for two PS examples of the Berlin test case after the stereo \ac{sar} processing. Step $1$ (gross outlier detection) and step $2$ ($2\sigma$ test) have already been applied.}
\label{fig:residual_plots}
\end{figure*}

%

\begin{figure*}[!tbh]
\centering
        \includegraphics[width=0.7\textwidth]{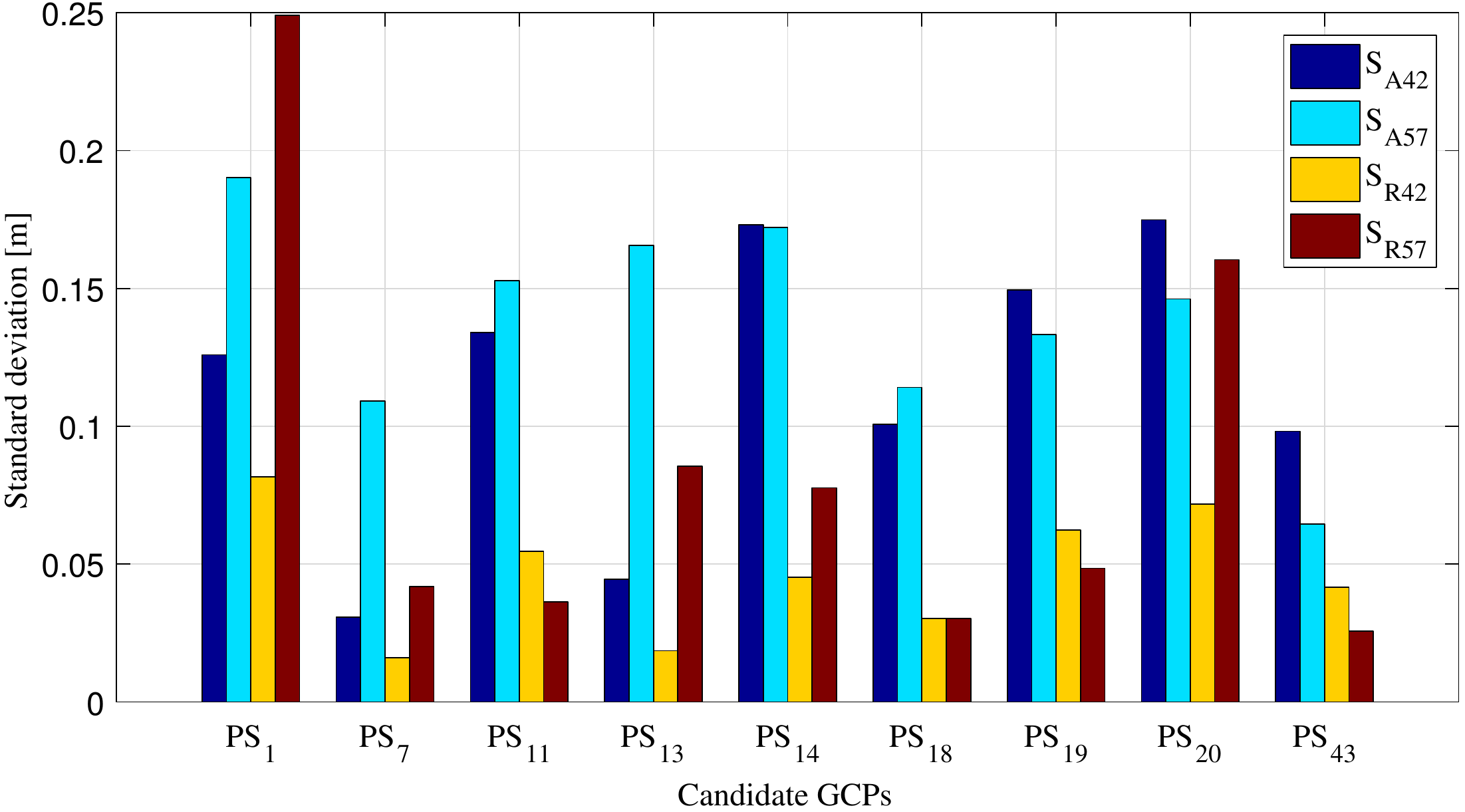}
        \caption{The posterior azimuth and range standard deviations ($S_{az}$, $S_{rg}$) of the 9 best \ac{gcp} candidates estimated by VCE in the geodetic stereo \ac{sar} processing.}\label{fig:bar_graph_obs}
\end{figure*}

\begin{figure*}[!tbh]
\centering
        \includegraphics[width=0.7\textwidth]{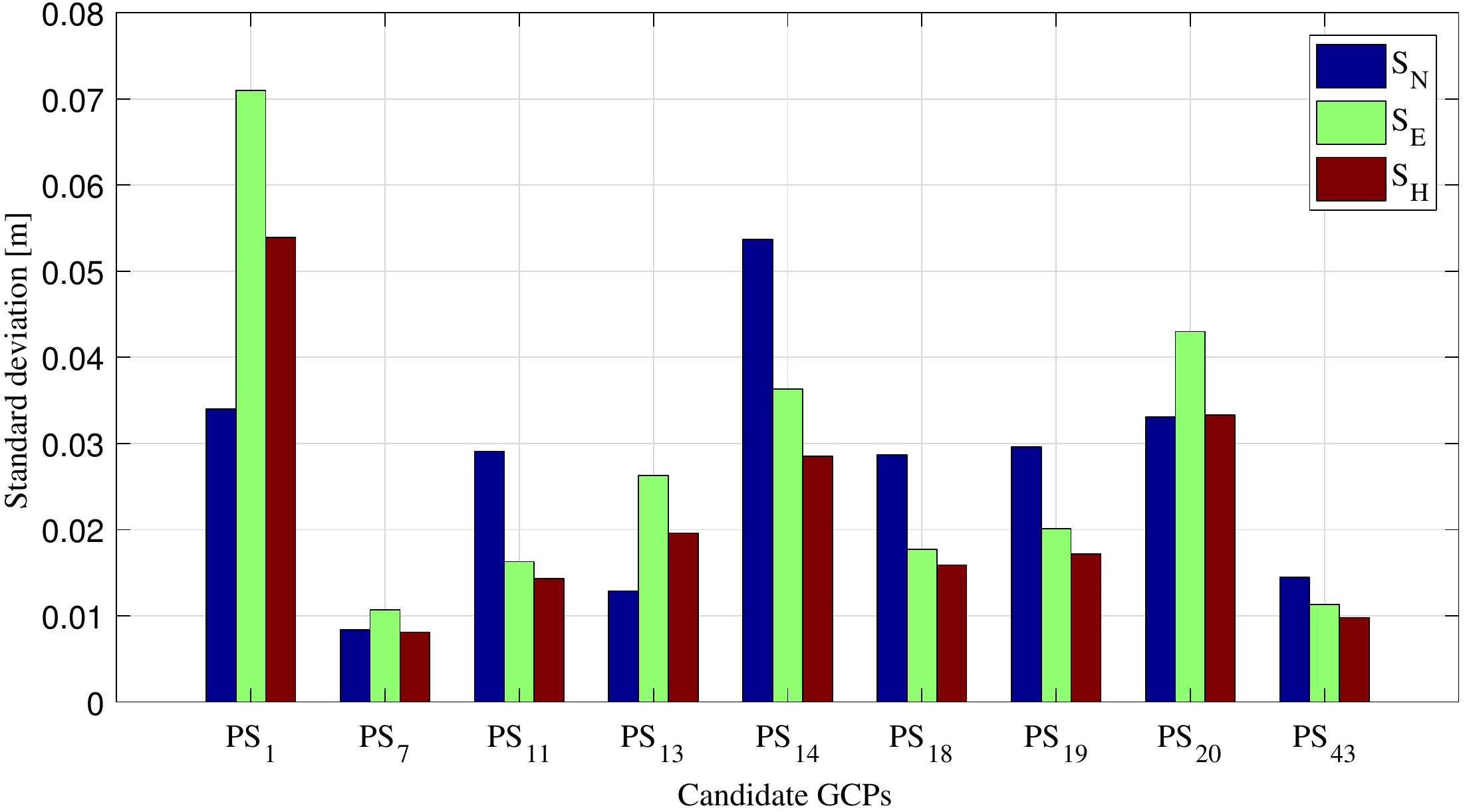}
        \caption{The posterior standard deviations scaled to 95$\%$ confidence level of the 9 best \ac{gcp} candidates estimated posterior to the geodetic stereo SAR processing. The standard deviations are defined in the local coordinate system of Berlin in north, east and vertical direction ($S_{N}$, $S_{E}$, $S_{H}$).}
\label{fig:bar_graph_coords}
\end{figure*}

The positioning of the 44 \ac{ps}s is carried out using stereo \ac{sar} (see Subsection \ref{ssec:backgr}) and is followed by outlier elimination according to criteria described in Subsection \ref{ssec:stereo_3D}. Starting from the corrected input timings, the first solution is analyzed for gross outliers in the observations exceeding the resolution of the underlying \ac{tsx} high-resolution spotlight product. Applying these thresholds to the residuals of the adjusted observations reduces the number of solvable \ac{ps}s from 44 to 42, because in the case of an obvious mismatch, all the observations of one geometry are removed. Moreover, the total number of observations is reduced by 25$\%$ but this strongly varies across the individual \ac{ps}s. After re-computation and application of the $2\sigma$ test using the estimated standard deviations from the VCE ($S_{az}$ , $S_{rg}$), the number of solvable \ac{ps}s remains 42 and the total amount of observations is reduced by another 8$\%$. At this stage, the data is fairly cleaned at the observation level regarding the individual range and azimuth geometries, but their consistency has not been considered so far. Looking at the observation residuals of the two \ac{ps}s displayed in Fig. \ref{fig:residual_plots} reveals that there are cases for which the azimuth of one geometry is clearly biased because we try to combine data from different phase centers. In the ideal case, the algorithm yields a coordinate solution for which all sets of observations (two sets of azimuth, two sets of range) can fulfill the range-Doppler positioning model of (\ref{eq:rg_eq}) and (\ref{eq:az_eq}). For a mismatch or a spatial separation of the phase centers, the usually more precise range observations dominate the solution, and only one of the azimuth data sets may fit the estimated coordinates without a bias, but not the second set of azimuth data. Such a situation is illustrated by $\text{PS}_{41}$ (see Fig. \ref{fig:residual_plots1}), where the $\text{36}^{\circ}$ azimuth displays a prominent bias of about -30 cm. To a certain degree this must be accepted since we can not expect ideal multi-directional \ac{ps}s, e.g. the lamp-poles have a certain diameter. In this study, as was mentioned in Subsection \ref{ssec:stereo_3D} we define an empirical limit of 20 cm of what we consider acceptable which removes candidates like the $\text{PS}_{41}$ during the final processing step. Therefore, the residual results of scatterers like $\text{PS}_{43}$ (see Fig. \ref{fig:residual_plots2}) may be seen as a best case scenario. The remaining difference in quality between range and azimuth is due to the non-square product resolution, i.e. the TS-X spotlight SLC data has a resolution of 0.6 m $\times$ 1.1 m in range and azimuth, respectively \cite{fritz_terrasar-x_2008}.\par


For the aforementioned reasons, only 9 \ac{ps}s remain which we consider as good \ac{gcp} candidates. The bar graphs of Fig. \ref{fig:bar_graph_obs} and Fig. \ref{fig:bar_graph_coords} summarize the quality of these  \ac{ps}s. Fig. \ref{fig:bar_graph_obs} shows the posterior estimated standard deviations of observations. The $S_{az}$ values vary from 3 cm to 19 cm with an average of 12 cm while the $S_{rg}$ values range from 1 cm to 24 cm with an average of 6 cm. This indicates that, for these natural \ac{ps}s, the removal of error terms, as expressed in (\ref{eq:rg_zeroD}) and (\ref{eq:az_zeroD}), and discarding the outliers allow sub-decimeter and decimeter precision in range and azimuth, respectively. In Fig. \ref{fig:bar_graph_coords}, the positioning quality is assessed by reporting the precision of the estimated coordinates. The standard deviations are defined in the local coordinate system of Berlin in north, east, and height ($S_{N}$, $S_{E}$, $S_{H}$) in the confidence level of 95$\%$. The mean values of $S_{N}$, $S_{E}$ and $S_{H}$ are 2.7 cm, 2.8 cm and 2.2 cm, respectively. The higher precision in the height direction is merely the effect of the cross-heading geometry used for the positioning.\par

Until now the discussion was mostly focused on analyzing the relative accuracy of the coordinates based on the posterior precision estimates. Although, the most reasonable procedure to validate the absolute accuracy is a point-wise comparison of stereo \ac{sar} coordinates with respect to the corresponding \ac{gnss}-surveyed ones, this was not applicable at the time of the study. Instead, the stereo \ac{sar} estimated ellipsoidal heights of the 9 \ac{gcp}s were compared with the LiDAR heights of the same area. We assume the phase centers of the detected \ac{gcp}s are at the base of the lamp poles on the ground. Therefore, the cross-comparison includes finding the nearest neighbors of the \ac{gcp} candidate in the LiDAR point cloud within the radius of 1 m, excluding the LiDAR points with large height values which originate from the top of the lamp pole, estimating the mode of the LiDAR heights to represent the reference height and evaluating the difference between the ellipsoidal height of stereo \ac{sar} results with respect to the reference height. The radius of the neighborhood is chosen in a way that the reference height calculation includes a reasonable number of samples and still be small enough to possibly prevent the inclusion of different objects in the search window. It is also worth to note that the calculation of mode is carried out with the assumption that the majority of samples in the window stem from the ground. The results of the cross-comparison are reported in Tab. \ref{tab:LiDAR_Comp}. The estimated stereo \ac{sar} and approximated LiDAR reference heights are denoted by ${\text{h}_{\text{S}}}$ and ${\text{h}_{\text{L}}}$ while their offset is represented by ${\text{h}_{\text{o}}}$. It is seen that for all except for one of the \ac{gcp}s the height offset is below 20 cm. The results report a bias of 13 cm and a precision of 5 cm overall with respect to the LiDAR data which roughly implies the absolute accuracy of the height estimation using the stereo \ac{sar} method.

\begin{table}[tbh!]
\centering
\ra{1.3}
\caption{\textsc{The result of cross-comparison between the estimated heights of stereo \ac{sar}} ${\text{h}_{\text{S}}}$ \textsc{and their corresponding LiDAR heights} ${\text{h}_{\text{L}}}$. \textsc{The offset} ${\text{h}_{\text{o}}}$ \textsc{is an indicator for the absolute accuracy of} ${\text{h}_{\text{S}}}$}
  \begin{tabular}{|l|ccc|}
    \hline
    \textbf{GCP} & ${\text{h}_{\text{L}}}~\text{[m]}$ & ${\text{h}_{\text{S}}}~\text{[m]}$ & ${\text{h}_{\text{o}}} = {\text{h}_{\text{S}}} - {\text{h}_{\text{L}}}~\text{[m]}$  \\
    \hline
    ${\text{PS}_{1}}$      &  74.64                  & 74.80                & 0.16  \\

    ${\text{PS}_{7}}$      & 74.45                   & 74.54                & 0.09  \\

    ${\text{PS}_{11}}$     & 74.83                   & 74.99                & 0.16  \\

    ${\text{PS}_{13}}$     & 75.40                   & 75.62                & 0.22  \\

    ${\text{PS}_{14}}$     & 73.87                   & 73.96                & 0.09  \\

    ${\text{PS}_{18}}$     & 73.87                   & 73.95                & 0.08  \\

    ${\text{PS}_{19}}$     & 75.59                   & 75.76                & 0.17  \\

    ${\text{PS}_{20}}$     & 75.02                   & 75.18                & 0.16  \\

    ${\text{PS}_{43}}$     & 79.78                   & 79.85                & 0.07  \\ \cline{1-4}

    Mean                   &                           &                        & $\mathbf{0.13}$  \\

    Standard deviation     &                           &                        & $\mathbf{0.05}$  \\ \cline{1-4}

  \end{tabular}
  \label{tab:LiDAR_Comp}
  \end{table}

\subsection{Oulu}
\label{ssec:oulu}

\begin{figure*}[tbh!]
\centering
\begin{subfigure}{0.49\textwidth}
\centering
\includegraphics[width =\textwidth]{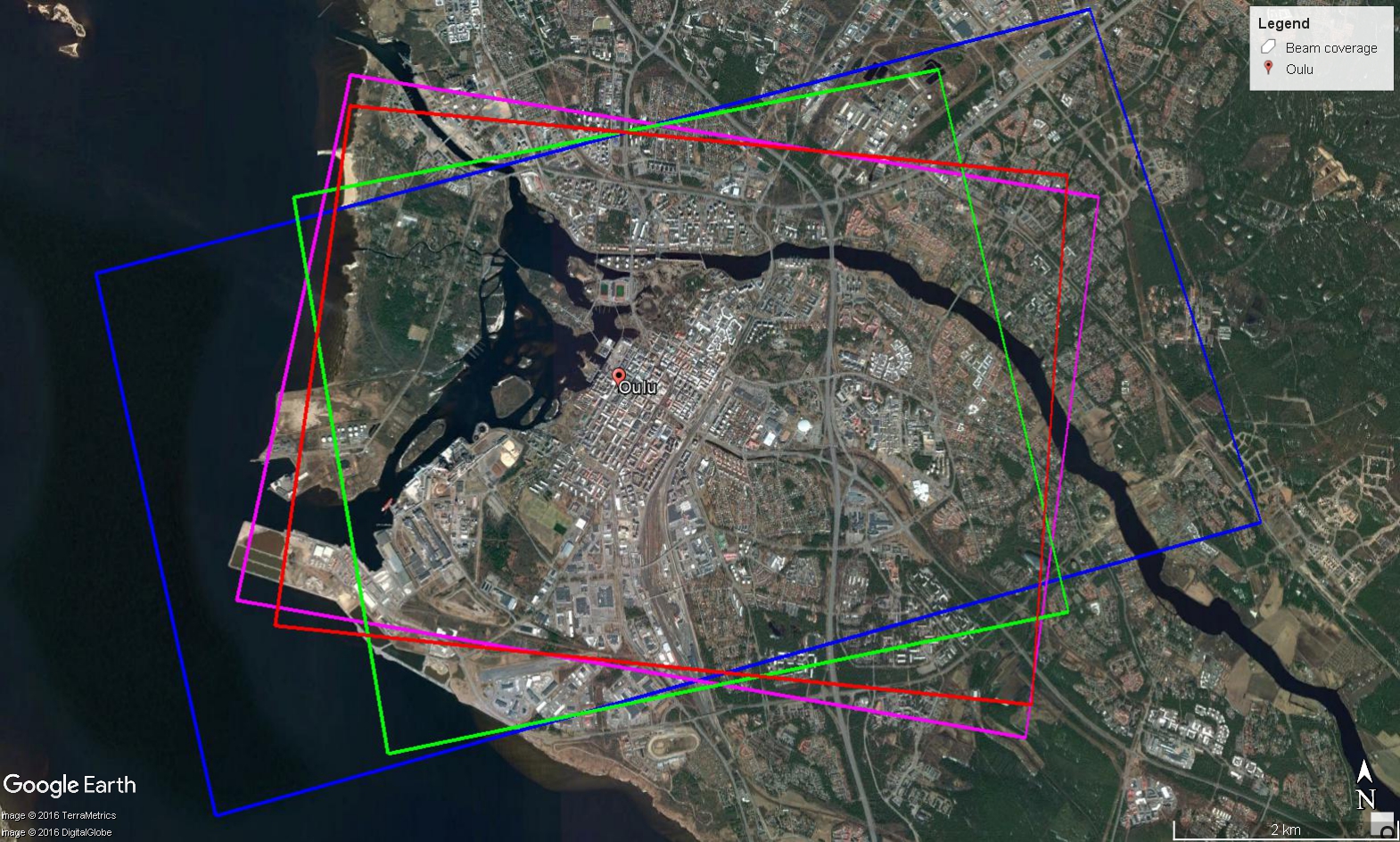}
\caption{}
\label{fig:scene_coverage_oulu}
\end{subfigure}
\begin{subfigure}{0.49\textwidth}
\centering
\includegraphics[width = \textwidth]{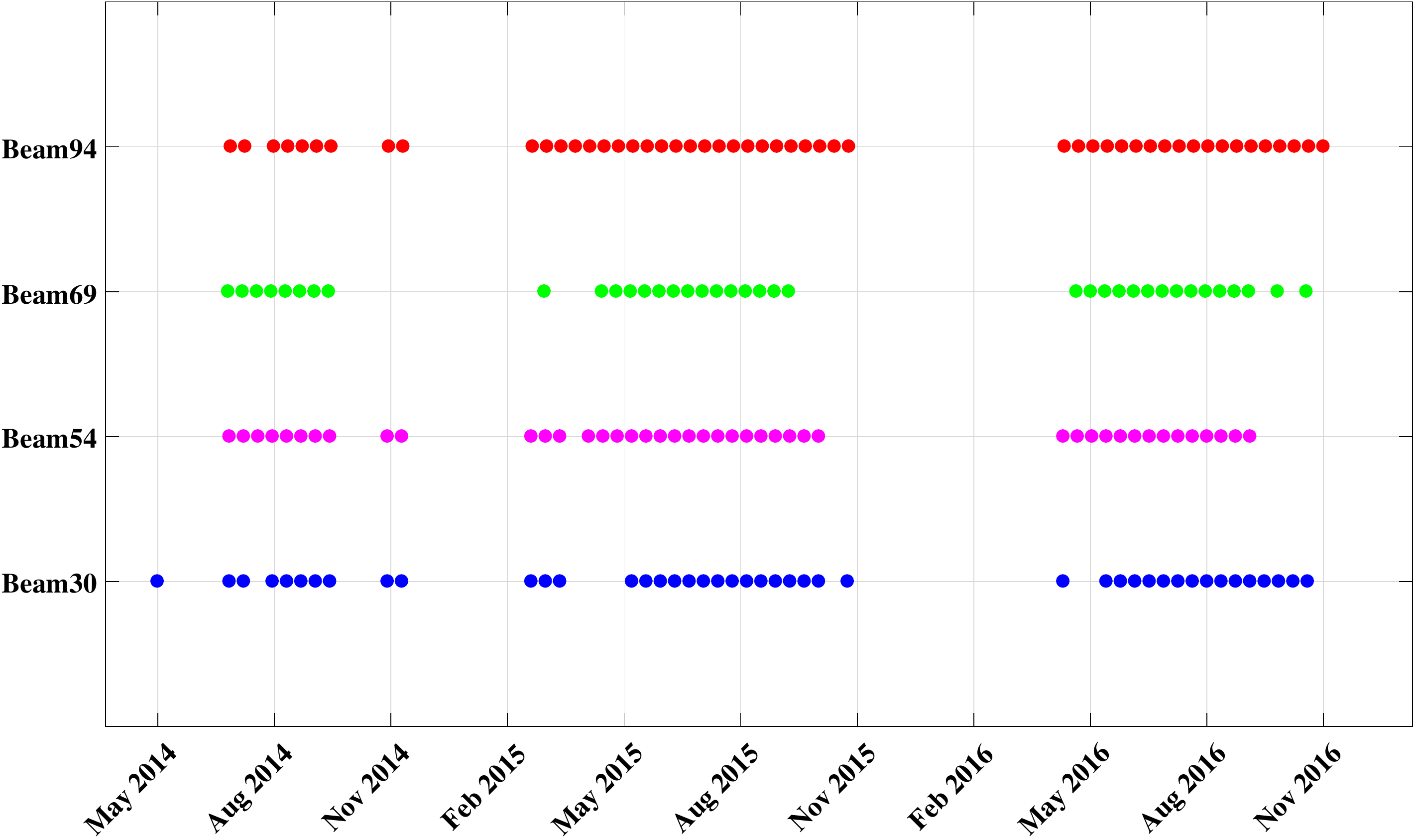}
\caption{}
\label{fig:acq_time_plot}
\end{subfigure}
\caption{(a) The mean scene coverage of the \ac{tsx} images overlaid on the optical image of Oulu taken from Google Earth. (b) The acquisition time plot of the \ac{tsx} images of Oulu.}
\label{fig:mean_scene_Oulu}
\end{figure*}

The second test site covers the entire city of Oulu. The \ac{sar} data include four stacks of \ac{tsx} high resolution spotlight products with 177 images in total. The images were acquired from May 2014 to October 2016, from two ascending orbits and two descending orbits. The acquisition parameters of the Oulu data set are reported in Tab. \ref{tab:acq_oulu} while the mean scene coverage and the acquisition time plot of the \ac{tsx} images are shown in Fig. \ref{fig:mean_scene_Oulu}. No images were ordered during the periods from November 2014 to March 2015 and November 2015 to March 2016 due to the accumulation of snow expected for Oulu during the winter months.

\begin{center}
\begin{table}[tbh!]
\centering
\ra{1.6}
\caption{\textsc{Acquisition parameters of stacks of \ac{sar} images in Oulu}}
  \begin{tabulary}{0.48\textwidth}{|C|CCCC|}
    \hline
    \textbf{Beam Nr.} & \textbf{Incidence angle (degree)} & \textbf{Heading angle (degree)} & \textbf{Track type} & \textbf{Nr. of images} \\
    \hline
    30     & 30.9                   & 346.1                & Asc  & 44          \\

    54     & 41.1                   & 191.4                & Dsc  & 44           \\

    69     & 46.2                   & 350.0                & Asc  & 38           \\

    94     & 53.4                   & 187.5                & Dsc  & 51           \\ \cline{1-5}
  \end{tabulary}
  \label{tab:acq_oulu}
  \end{table}
\end{center}

For Oulu, no optical images with sufficient spatial resolution were available for the detection of \ac{ps} candidates from cross-heading geometries. Therefore, the road network data of Oulu was used instead. The data was freely accessed from the Finnish Transport Agency \cite{noauthor_finnish_2017}. It was delivered in vector format in the \ac{utm} coordinate system and includes the main streets and highways of Oulu.\par

\begin{figure*}[!tbh]
\centering
        \includegraphics[width=0.8\textwidth]{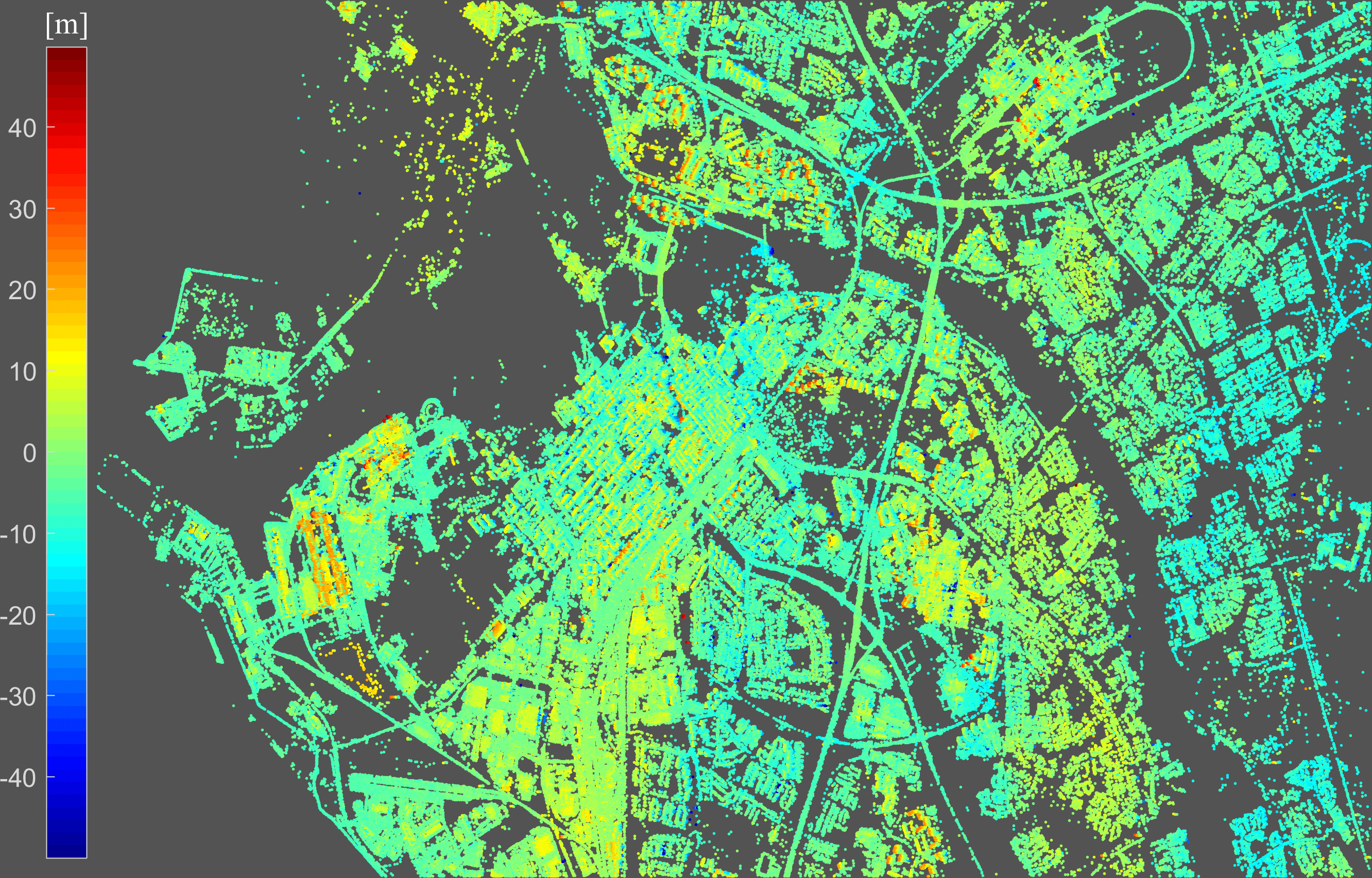}
        \caption{\ac{ps} elevation map obtained from \ac{psi} processing of an ascending stack of Oulu (beam 30). The total number of scatterers is approximately 540000 after selecting only the \ac{ps}s with posterior coherence values equal or higher than 0.7.}\label{fig:ps_pc_b30}
\end{figure*}

\begin{figure*}[tbh!]
\centering
\begin{subfigure}{1\textwidth}
\centering
\includegraphics[width = 0.8\textwidth, height=10cm]{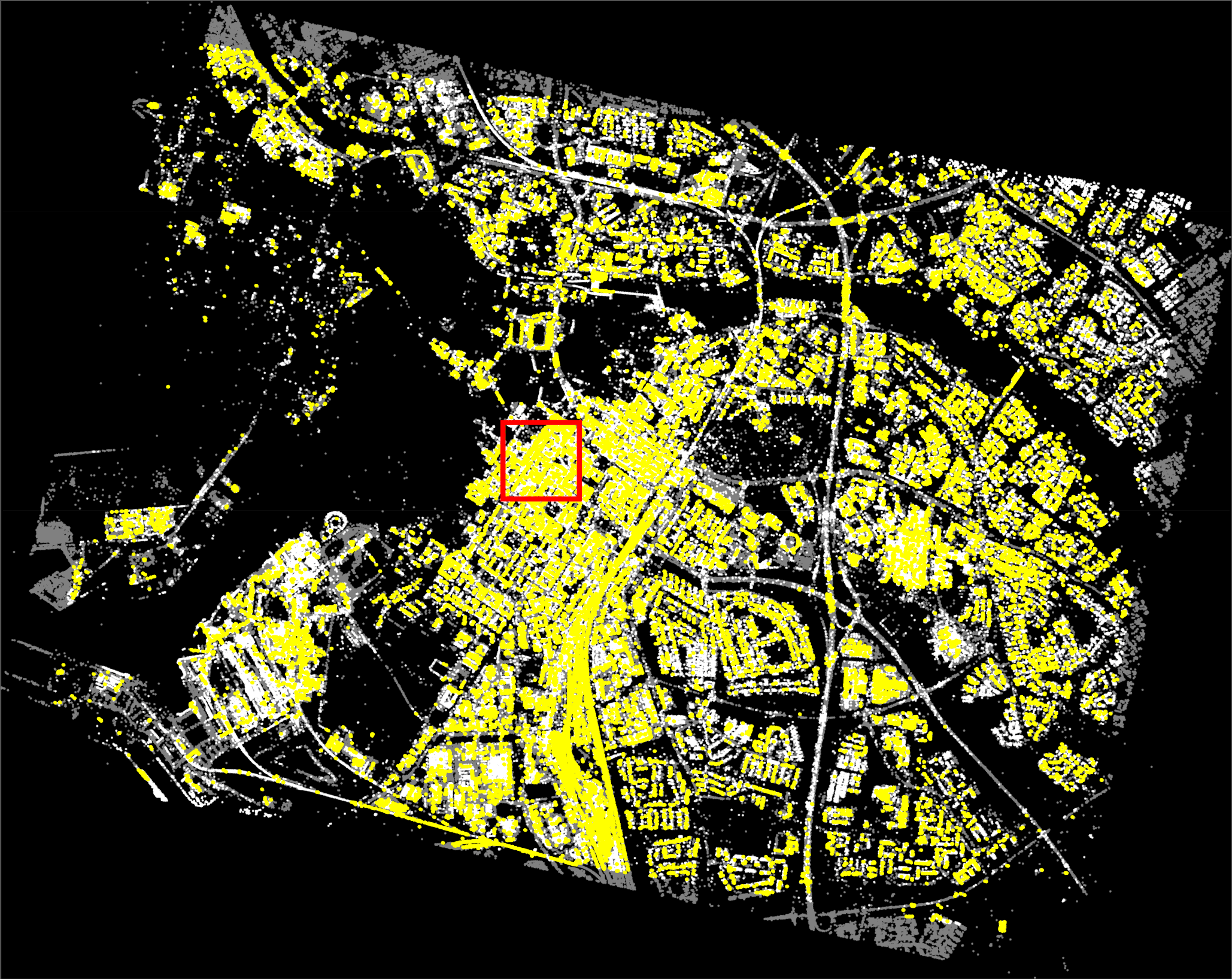}
\caption{}
\label{fig:Coarse_SH_matching_full}
\end{subfigure}
\begin{subfigure}{0.35\textwidth}
\centering
\includegraphics[width = \textwidth]{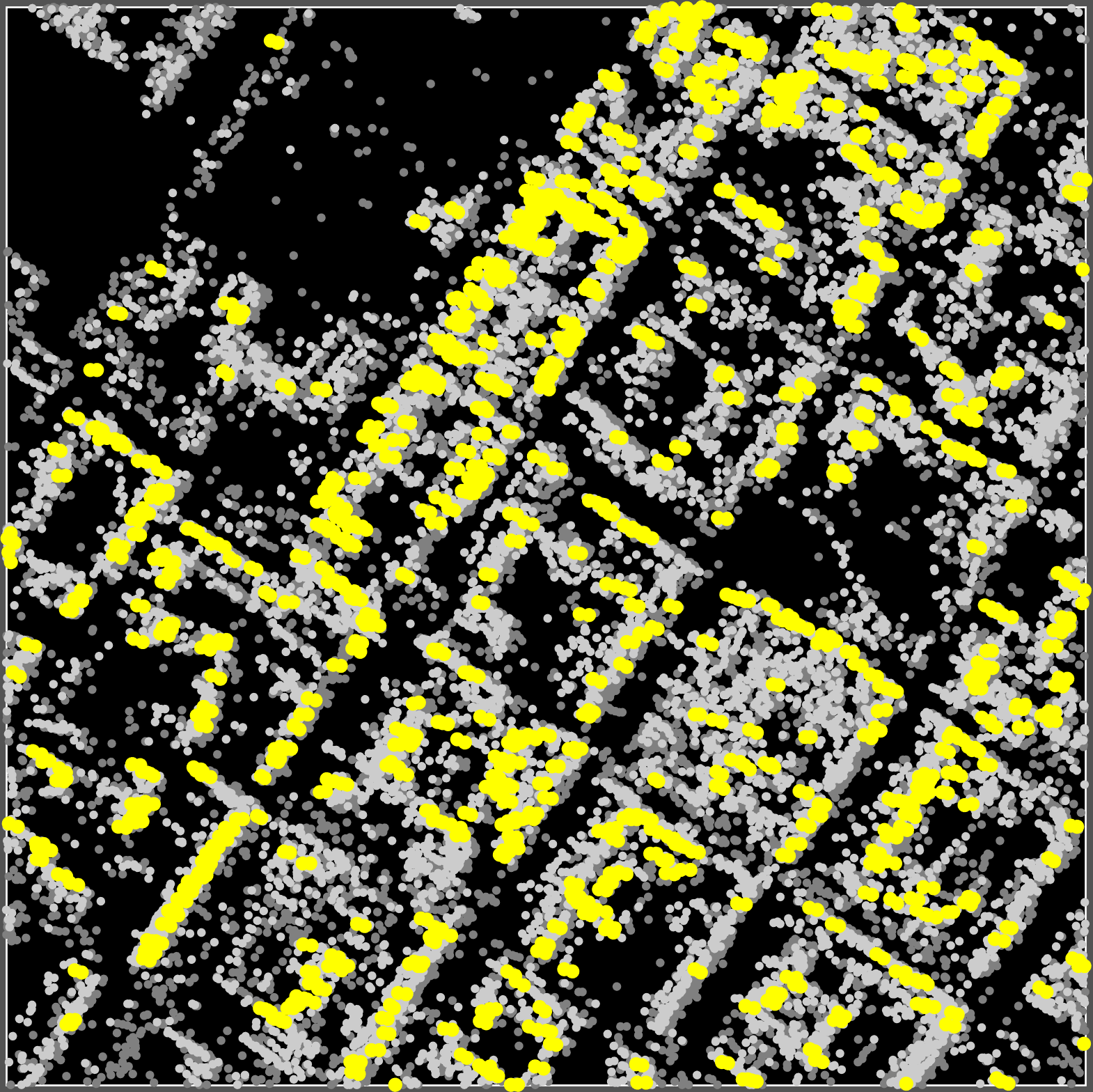}
\caption{}
\label{fig:Coarse_SH_matching_bef_red}
\end{subfigure}\hspace*{2cm}
\begin{subfigure}{0.35\textwidth}
\centering
\includegraphics[width = \textwidth]{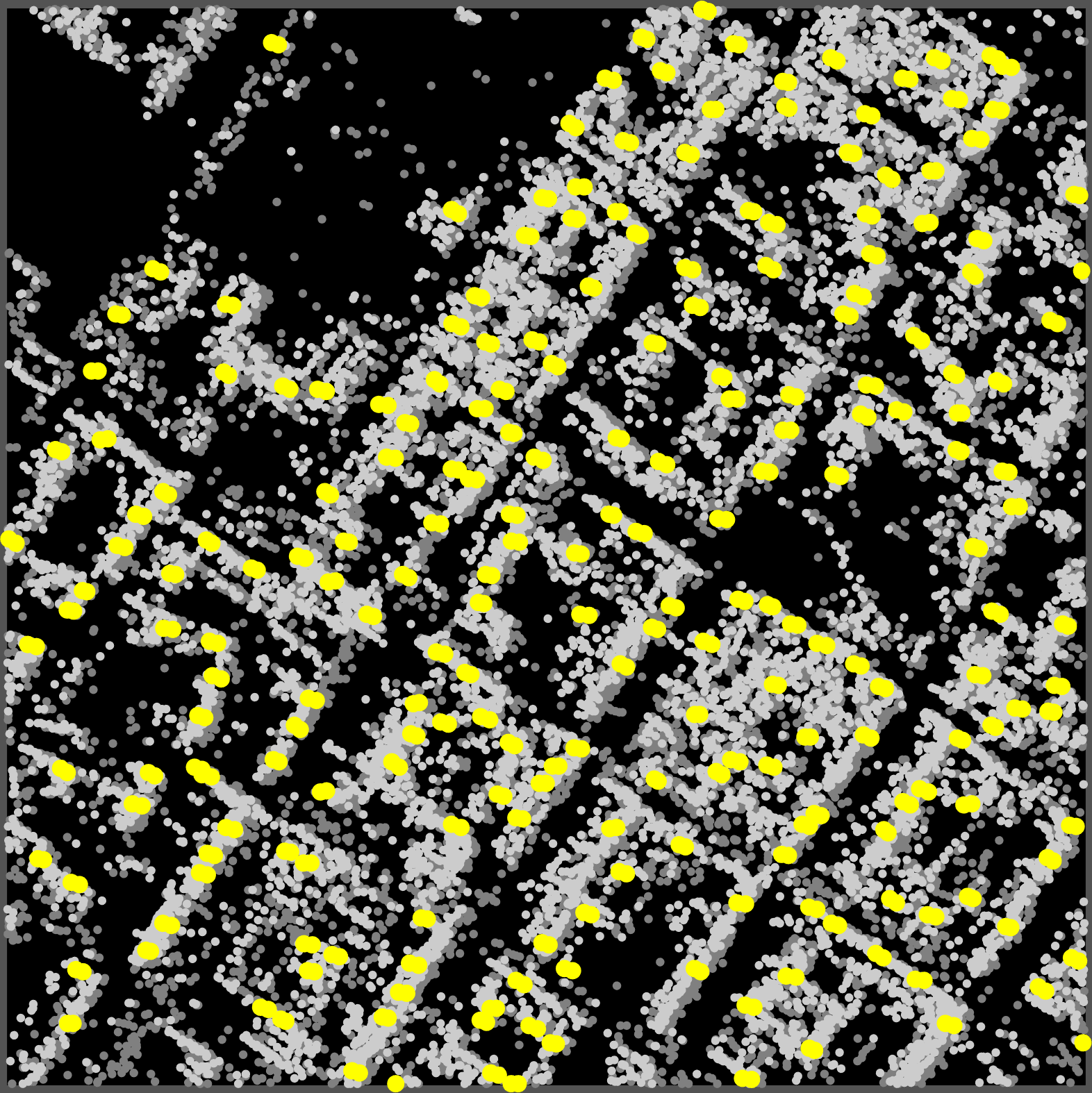}
\caption{}
\label{fig:Coarse_SH_matching_aft_red}
\end{subfigure}
\caption{Depiction of \ac{ps} correspondence detection from \ac{sar} images of same-heading orbit tracks of Oulu using the first step of the multitrack \ac{ps} fusion algorithm proposed in \cite{gernhardt_geometrical_2012}. (a) shows the geocoded \ac{ps} point clouds of beams 54 and 94 in a DD configuration as white and gray points as well as the detected \ac{ps} correspondences in yellow. (b) and (c) show a zoom-in area of (a) marked by the red rectangle before and after imposing a 10 m grid in which the pairs with closest distance are chosen.}
\label{fig:Oulu_SH_Detection}
\end{figure*}

The detection of \ac{ps} candidates from the same-heading tracks was carried out using the multitrack \ac{psi} fusion algorithm described in Subsection \ref{ssec:fusion}. As the prerequisite of the algorithm, the \ac{psi} processing was performed by the PSI-GENESIS of the German Aerospace Center (DLR) \cite{adam_wide_2013}. For each detected \ac{ps}, the elevation and the deformation parameters (in this study only a linear trend) were estimated. As an example, the radar-coded \ac{ps} elevation map of the ascending stack of beam 30 is visualized in Fig. \ref{fig:ps_pc_b30}. After geocoding, the \ac{3D} point clouds obtained from either ascending-ascending (AA) or descending-descending (DD) geometries form the initial input of the fusion algorithm as Fig. \ref{fig:Oulu_SH_Detection} demonstrates.

Fig. \ref{fig:Coarse_SH_matching_full} shows the geocoded \ac{ps} point clouds from the DD geometries, visualized in white and gray. The yellow points represent the identified \ac{ps} pairs from the fusion algorithm. The total number of the correspondences is approximately 32000 and the Euclidean distances between the matched \ac{ps}s vary from 1.5 to 5 meters. In order to reduce the number of \ac{ps} correspondences to the ones with higher quality and closer distance, as well as to preserve the homogeneity of the distribution, a regular grid was imposed on the point clouds. Inside the 10 m $\times$ 10 m grid, the \ac{ps} pairs which were closer together and had lower \ac{adi} values were selected to reduce the number of pairs from 32000 to 10000. The comparison between \ac{ps} pairs before and after reduction can be seen in a zoomed-in area in Fig. \ref{fig:Coarse_SH_matching_bef_red} and Fig. \ref{fig:Coarse_SH_matching_aft_red}, respectively. The same procedure depicted in Fig. \ref{fig:Oulu_SH_Detection} was also carried out for the \ac{sar} images from the AA geometries and close to 9500 \ac{ps} correspondences were detected. The results then were radar-coded for both geometry configurations.

\begin{figure*}[tbh!]
\centering
\begin{subfigure}{0.42\textwidth}
\centering
\includegraphics[width = \textwidth]{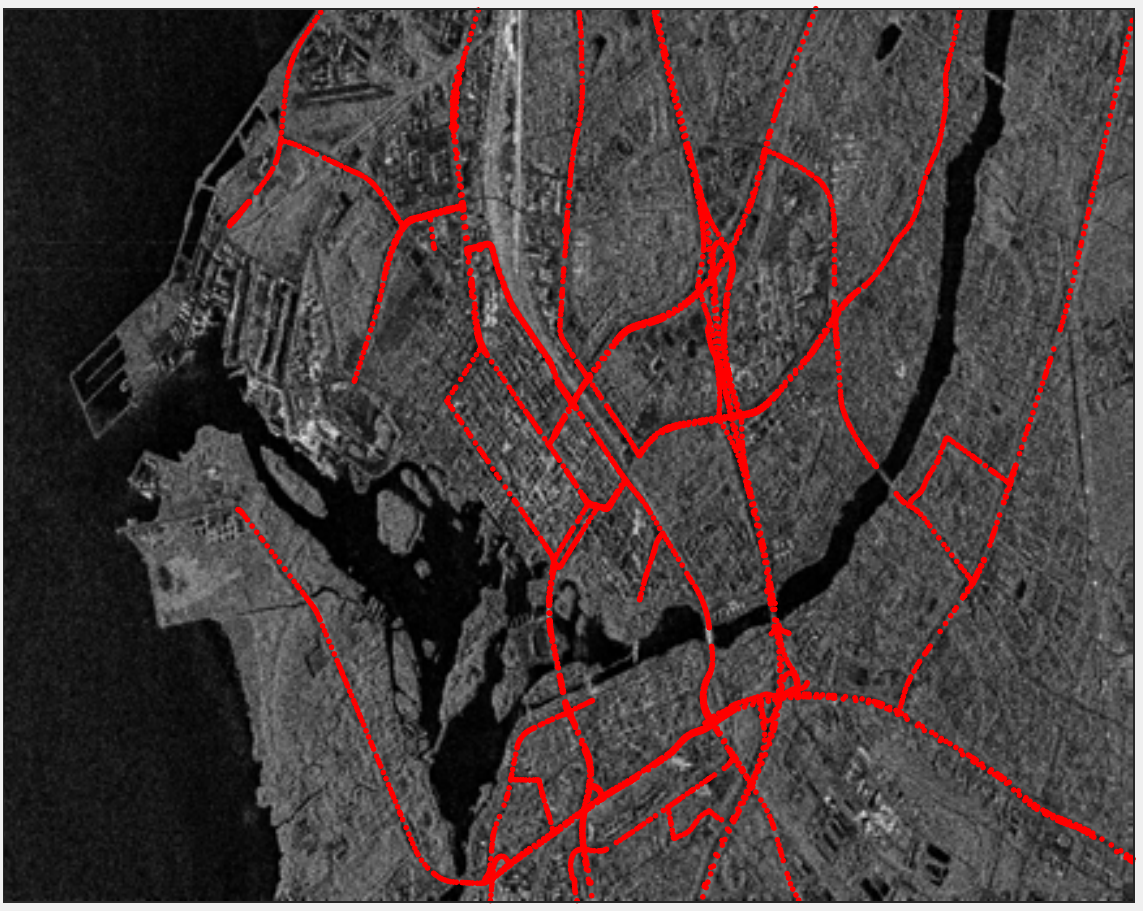}
\caption{}
\label{fig:road_30}
\end{subfigure}\hspace*{.35cm}
\begin{subfigure}{0.42\textwidth}
\centering
\includegraphics[width = \textwidth]{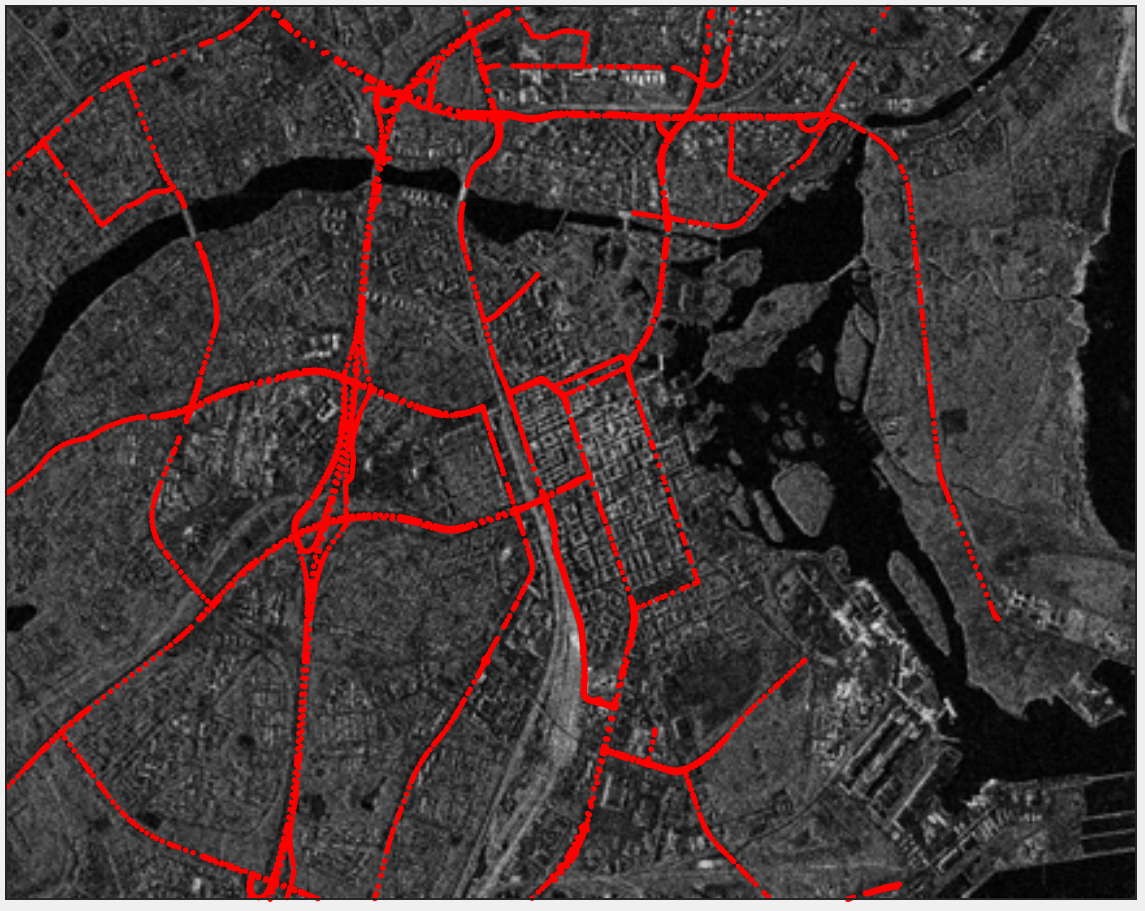}
\caption{}
\label{fig:road_54}
\end{subfigure}
\begin{subfigure}{0.42\textwidth}
\centering
\includegraphics[width = \textwidth]{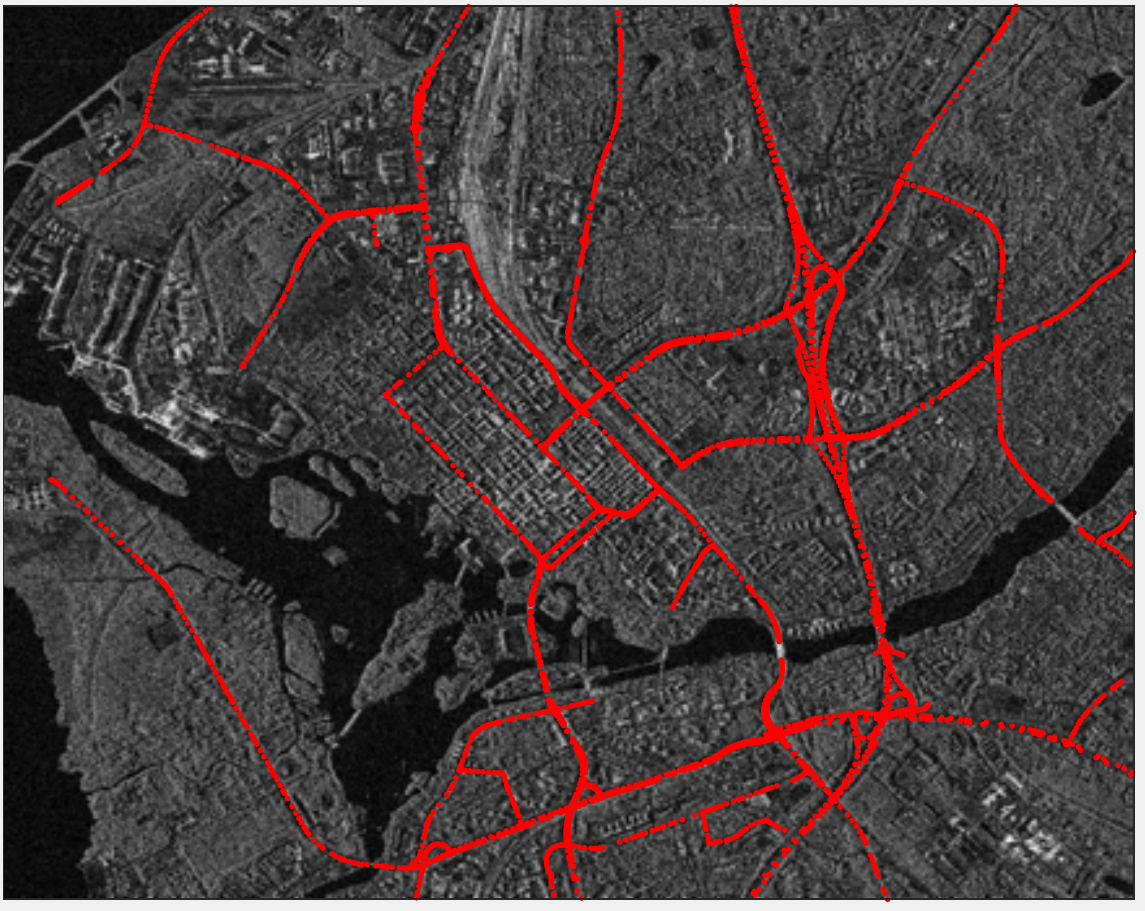}
\caption{}
\label{fig:road_69}
\end{subfigure}\hspace*{.35cm}
\begin{subfigure}{0.42\textwidth}
\centering
\includegraphics[width = \textwidth]{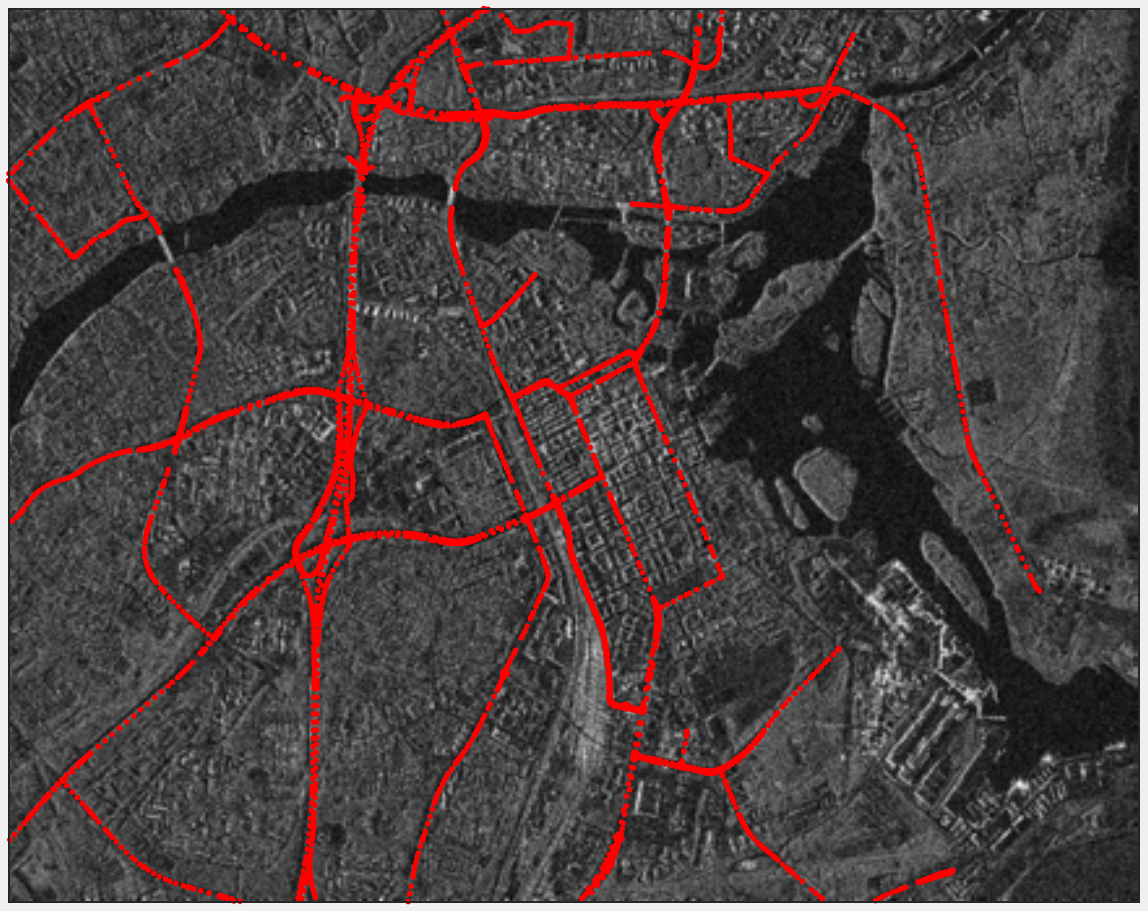}
\caption{}
\label{fig:road_94}
\end{subfigure}
\caption{Projection of the road network data of Oulu onto the master scene of (a) beam 30, (b) beam 54, (c) beam 69 and (d) beam 94. The road data is represented by red points and is the basis for detection of identical \ac{ps} candidates from cross-heading orbit geometries.}
\label{fig:road_vis}
\end{figure*}

\begin{figure*}[tbh!]
\centering
\begin{subfigure}{0.37\textwidth}
\centering
\includegraphics[width = \textwidth]{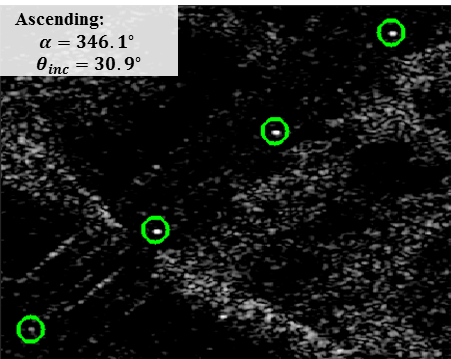}
\caption{}
\label{fig:ADAD30}
\end{subfigure}
\begin{subfigure}{0.37\textwidth}
\centering
\includegraphics[width = \textwidth]{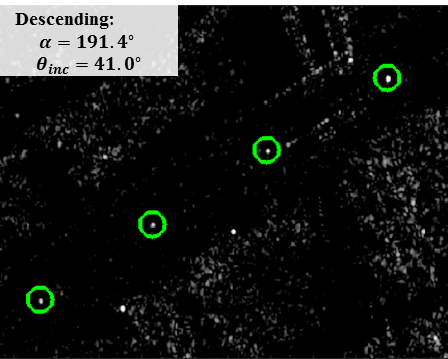}
\caption{}
\label{fig:ADAD54}
\end{subfigure}
\begin{subfigure}{0.37\textwidth}
\centering
\includegraphics[width = \textwidth]{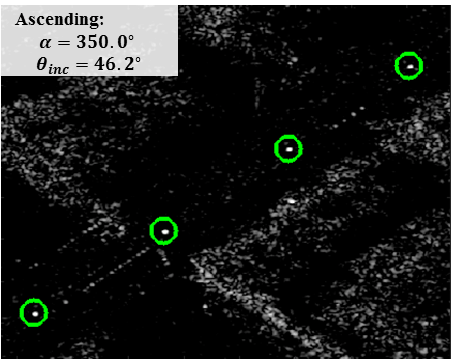}
\caption{}
\label{fig:ADAD69}
\end{subfigure}
\begin{subfigure}{0.37\textwidth}
\centering
\includegraphics[width = \textwidth]{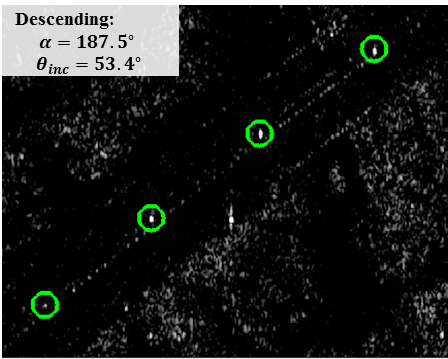}
\caption{}
\label{fig:ADAD94}
\end{subfigure}
\caption{\ac{ps} correspondence detection from ADAD geometry configuration. For each image, the respective averaged heading angle $\alpha$ and the averaged incidence angle $\theta_{inc}$ are stated.}
\label{fig:ADAD_candidates}
\end{figure*}

The \ac{ps} candidates to be localized from the cross-heading geometries, either ascending-descending (AD) configuration or quad geometry (ADAD) configuration, were selected based on the detection of bright points along the roads using the road network data as was explained in Subsection \ref{ssec:road_network}. The road network data was first radar-coded on the master scenes of all the four geometries as is seen on the mean intensity images of each stack in Fig. \ref{fig:road_vis}. A circular neighborhood with a radius of 70 pixels was then considered around each road network node within which the \ac{adi} was evaluated for all pixels for all the four stacks. The neighborhood is chosen based on a rough knowledge on the maximum width of highways in Oulu ($\approx 35$ m) and was adapted to the \ac{sar} data by taking into account the pixel spacing in range and azimuth direction and the oversampling factor used in the processing. After selecting the pixel with the lowest \ac{adi} in each neighborhood, a further threshold of 0.25 on \ac{adi} values, typically used in \ac{psi} processing \cite{ferretti_permanent_2001}, selects the stable bright points from each stack. After geocoding, the \ac{ps}s from different stacks which are closer than a threshold of three meters are chosen as the final stereo candidates and are subsequently radar-coded on all the \ac{sar} images. The distance threshold depends on the geometry configuration from which the user is interested to localize the targets. If same-heading geometries are considered the value should be lower than three meters in order to ensure correct \ac{ps} correspondence detection. An example of \ac{ps} candidates visible from ADAD configuration in Oulu is given in Fig. \ref{fig:ADAD_candidates}. The candidates are all assumed to be bases of lamp poles and can be seen as bright points inside the green circles. The explained procedure produced 107 and 52 initial candidates from ADAD and AD geometry configurations, respectively. The quantity is lower in the latter because of the strict distance threshold of 1.5 m imposed on coordinate differences of \ac{ps}s visible in different stacks. The threshold value is chosen empirically based on the histogram of minimum Euclidean distances evaluated between the PS pairs.\par

\begin{figure*}[t!] 
\begin{subfigure}{0.48\textwidth}
\includegraphics[width=\linewidth]{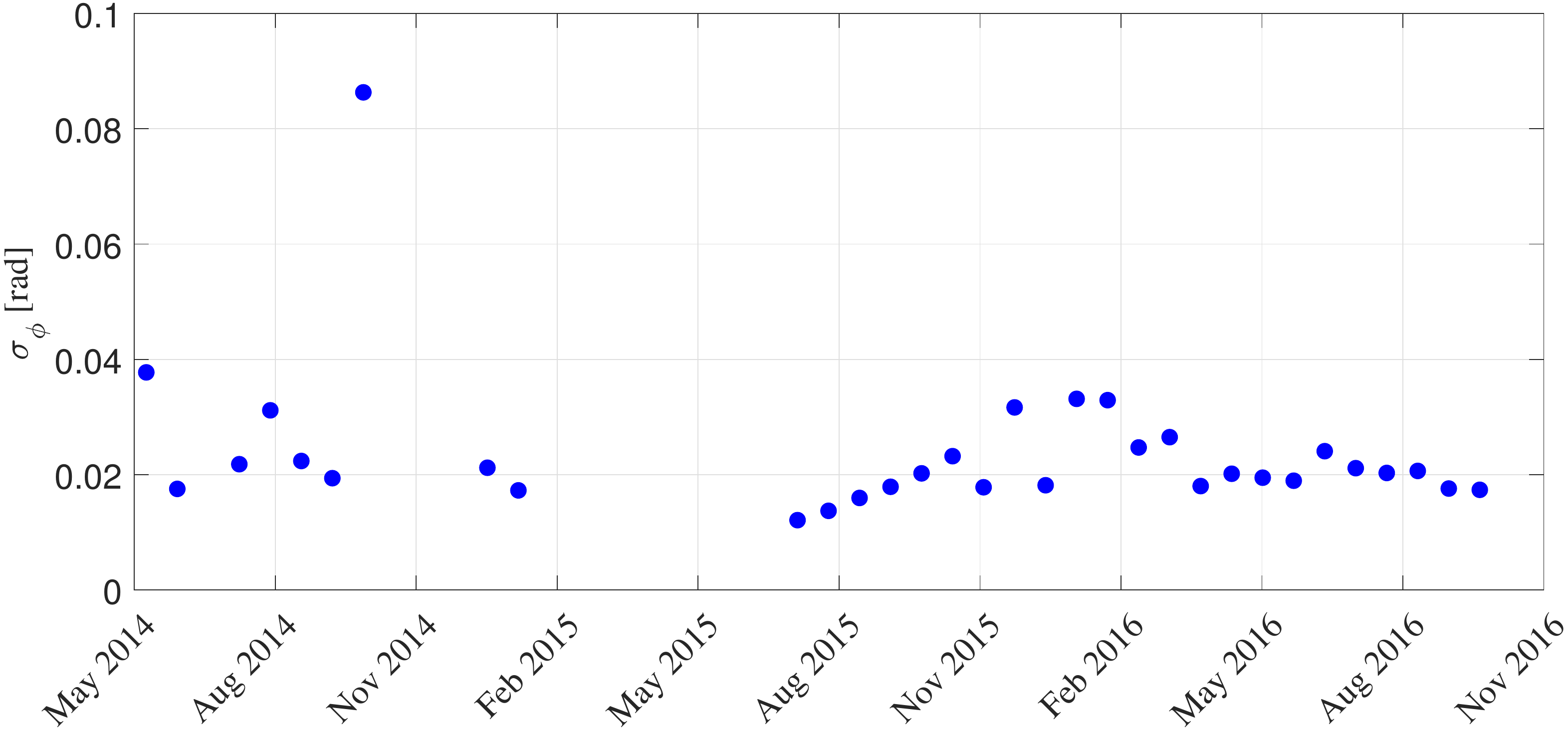}
\caption{}
\end{subfigure}\hspace*{\fill}
\begin{subfigure}{0.48\textwidth}
\includegraphics[width=\linewidth]{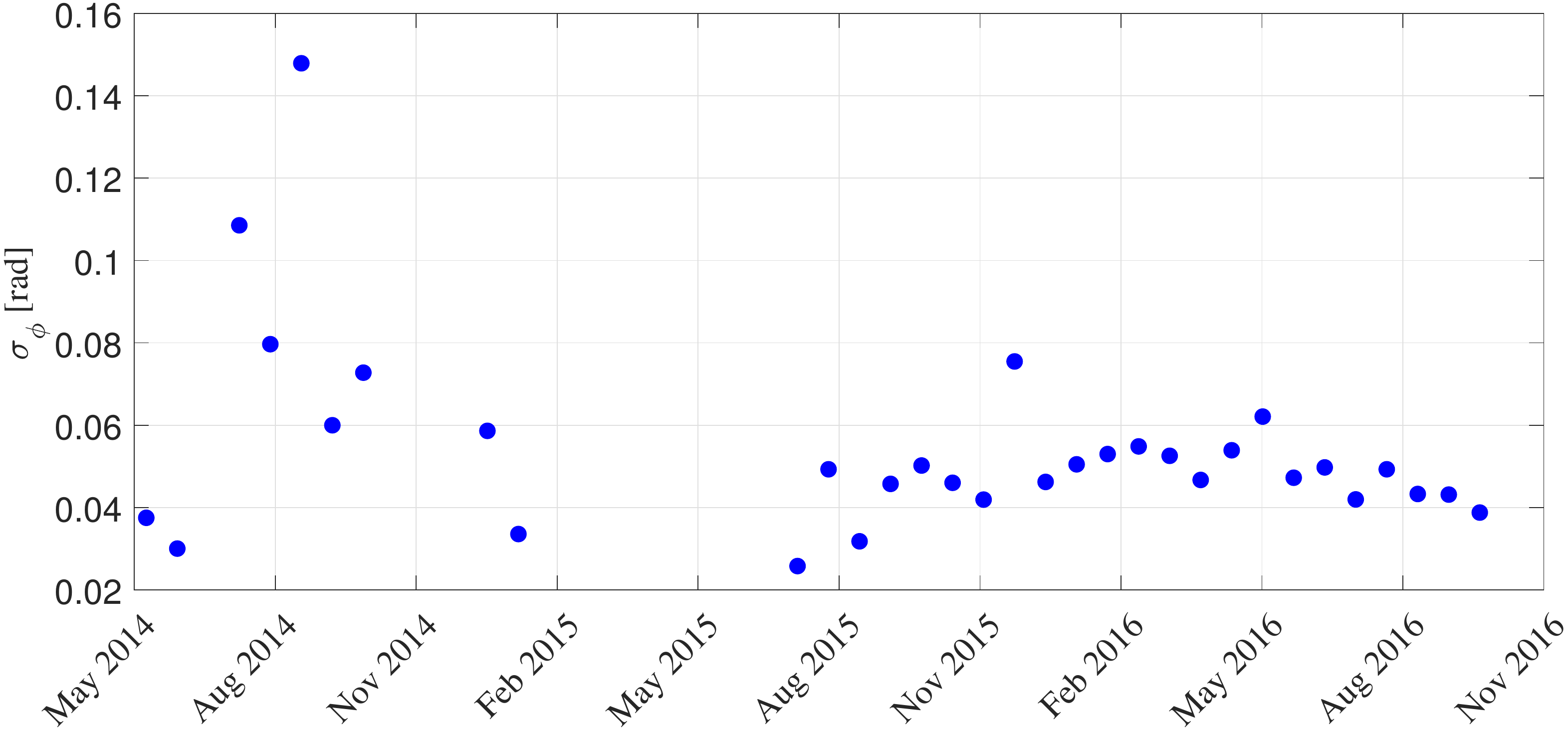}
\caption{}
\end{subfigure}

\medskip
\begin{subfigure}{0.48\textwidth}
\includegraphics[width=\linewidth]{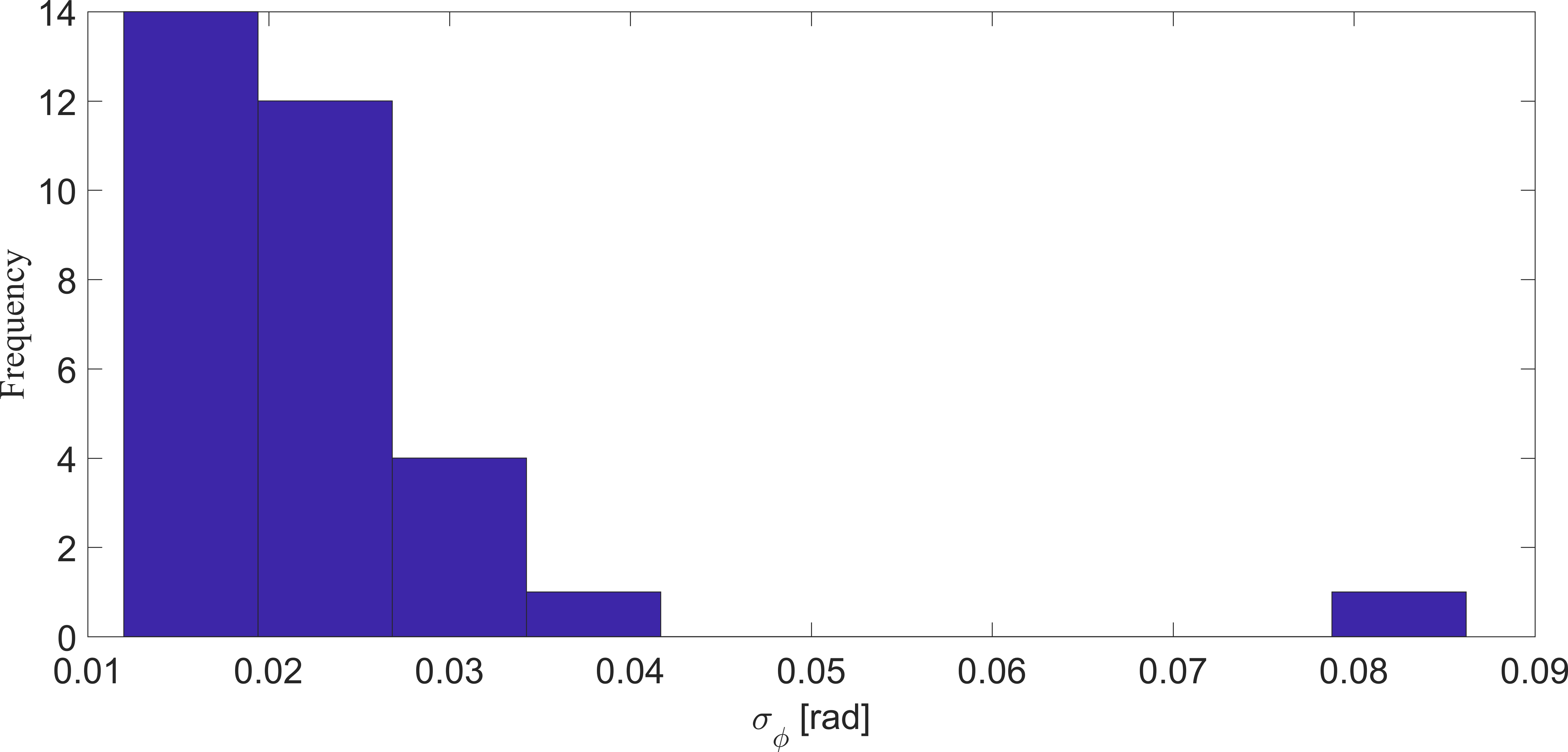}
\caption{}
\end{subfigure}\hspace*{\fill}
\begin{subfigure}{0.48\textwidth}
\includegraphics[width=\linewidth]{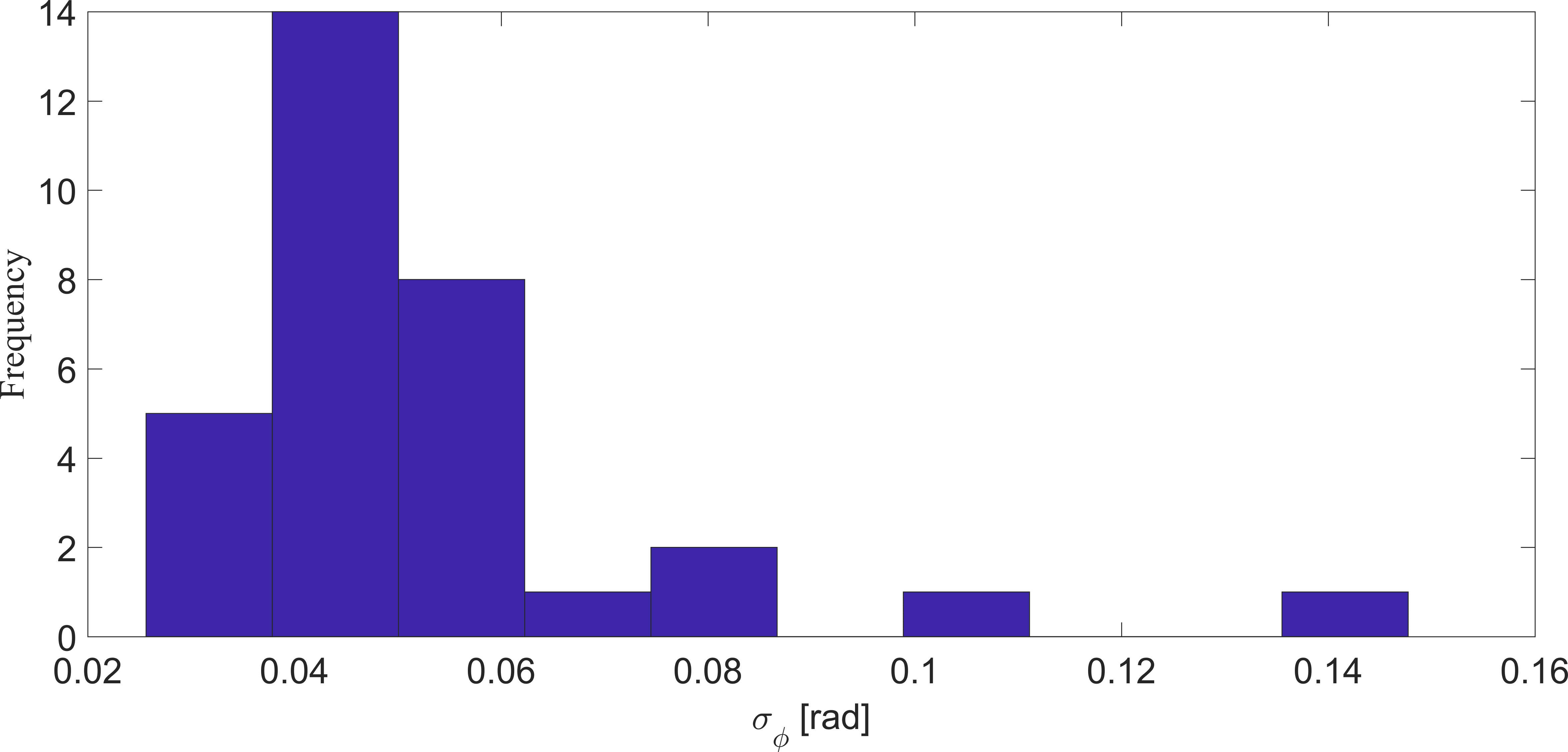}
\caption{}
\end{subfigure}

\medskip
\begin{subfigure}{0.48\textwidth}
\includegraphics[width=\linewidth]{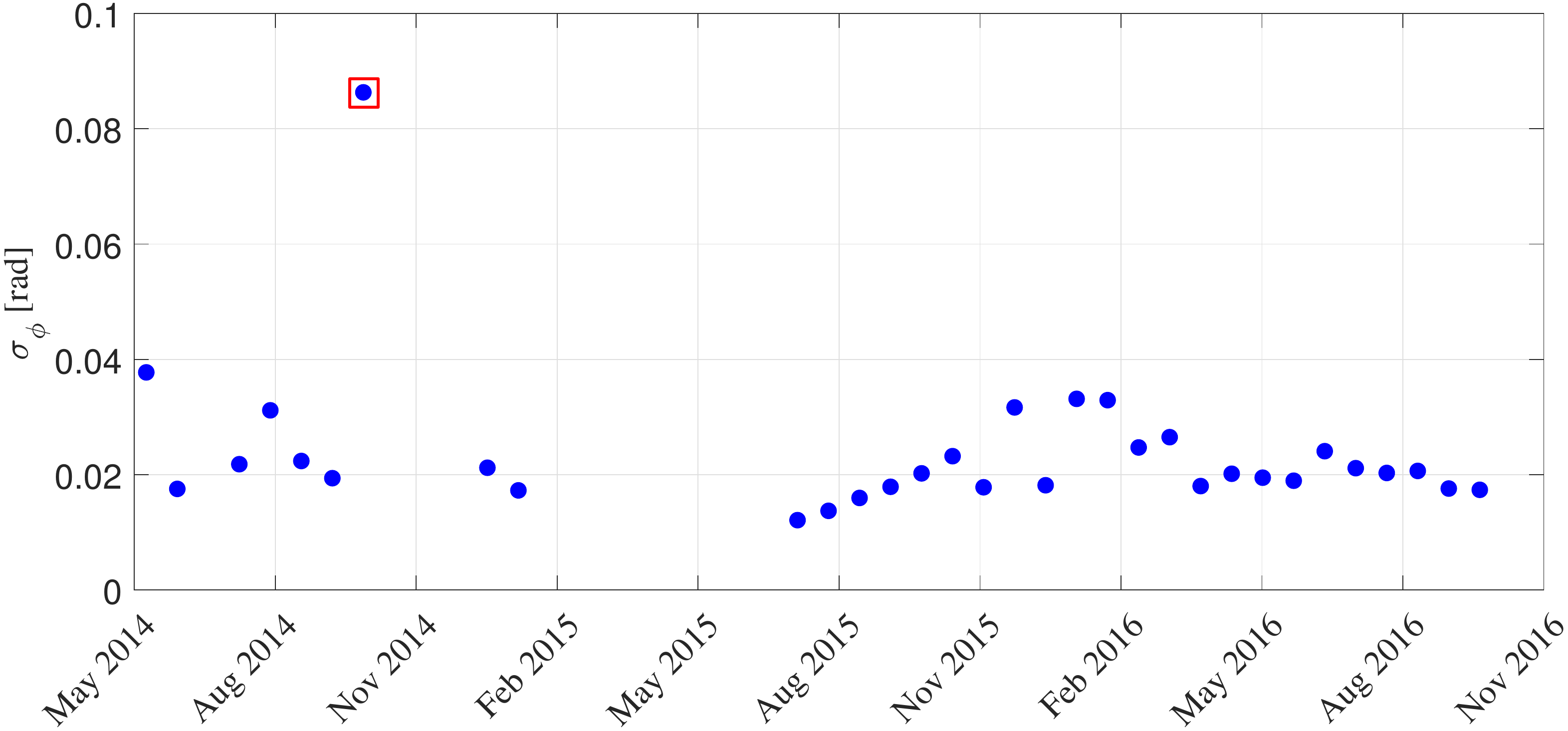}
\caption{}
\end{subfigure}\hspace*{\fill}
\begin{subfigure}{0.48\textwidth}
\includegraphics[width=\linewidth]{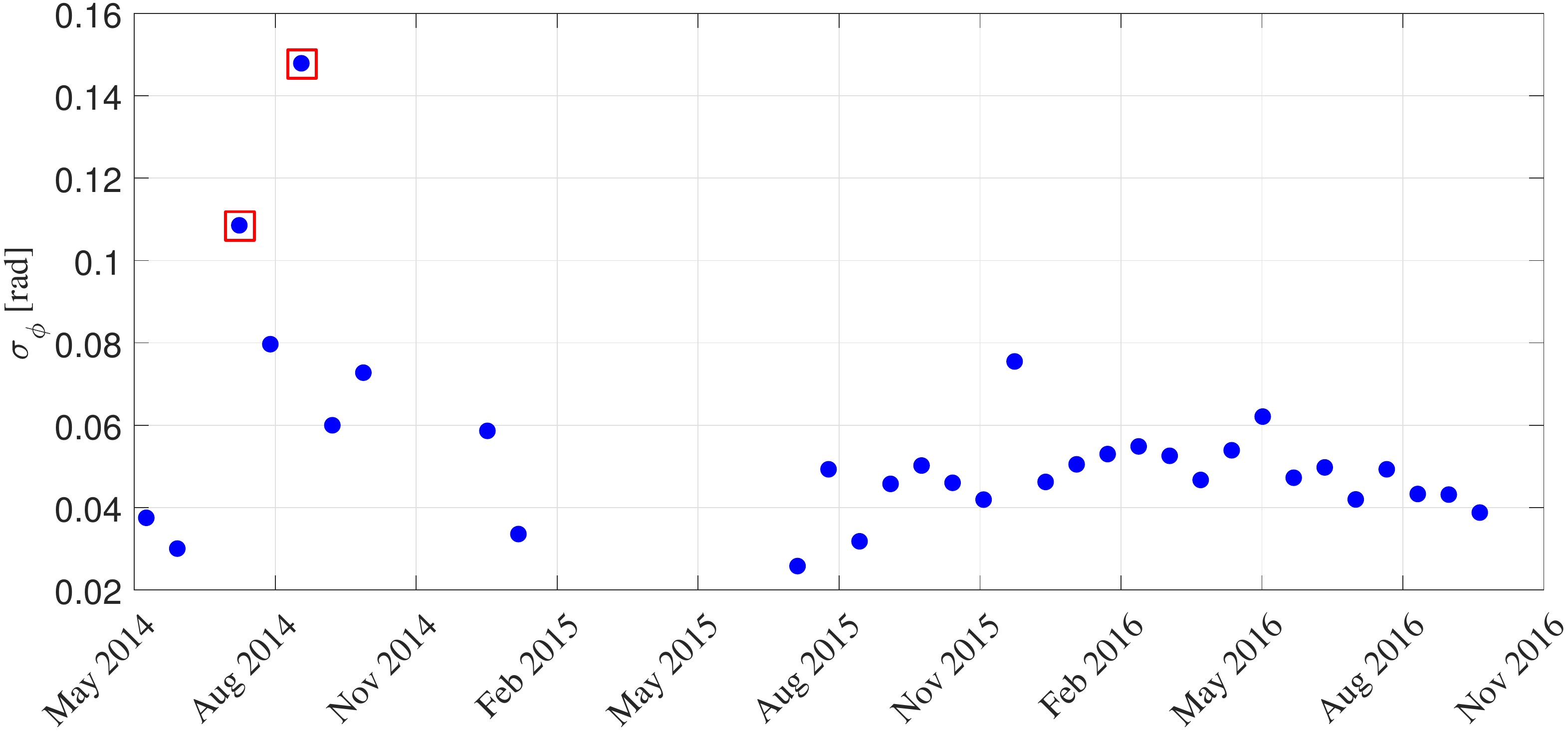}
\caption{}
\end{subfigure}

\caption{Example of initial outlier removal based on phase noise time series of the detected candidates. The outliers are identified and removed automatically based on the distribution of $\bm{\sigma_{\phi}}$.} \label{fig:bp_examples}
\end{figure*}

The \ac{pta} was performed on all the detected \ac{ps} candidates in all the \ac{sar} images in which the candidate was visible. For each candidate a time-series of $\bm{\sigma_{\phi}}$ was evaluated using (\ref{eq:phase_nouise}). Fig. \ref{fig:bp_examples} demonstrates the initial outlier removal on two \ac{ps} candidates based on the method of adjusted boxplot explained in Subsection \ref{ssec:outlier_rem}. The $\sigma_{\phi}$ values of each scatterer are sorted in time in Fig.13a and Fig.13b. The distribution of $\bm{\sigma_{\phi}}$ is right-skewed in both cases as can be seen in Fig.13c and Fig.13d. From the distribution plots, one can  detect the samples which are not connected to the tail of the distribution and mark them as potential outliers. This process is done automatically and without the need for manual intervention using the interval defined in (\ref{eq:bp_ad}). In Fig.13e and Fig.13f, the detected outliers are marked with red rectangles. For all the candidates the mentioned procedure was carried out.\par
%


The atmospheric corrections were carried out using global ionospheric maps and the global tropospheric zenith path delays provided with the Vienna mapping function \cite{boehm_troposphere_2006} since there was no access to the Oulu \ac{gnss} receiver at the time of the study. The geodynamic effects were fully considered according to the IERS conventions and all the effects were removed from the \ac{ps} timings.\par

The final positioning was carried out by stereo \ac{sar} with all the mentioned outlier removal steps in Subsection \ref{ssec:stereo_3D}. The last criterion, which includes the removal of \ac{ps}s based on $S_{az}$ values higher than 20 cm, reduces the amount of total \ac{ps}s to only those points which can be considered ideal stereo candidates. The averaged quality of the estimated \ac{3D} coordinates for all the remaining high quality \ac{ps}s are reported in Tab. \ref{tab:Oulu_res}. The scatterers are categorized based on the geometry configuration that is used for their positioning. The results are all expressed in centimeter and are defined in the local east, north and vertical coordinates. The standard deviations are all defined within 95$\%$ confidence interval. From Tab. \ref{tab:Oulu_res}, it is seen that the averaged precisions are smaller than two decimeter for all the cases. As it was expected, the localization quality boosts as the difference in the viewing geometries becomes larger which is the case when changing from AA or DD to the AD and ADAD geometry configurations. It is also evident that for cross-heading geometries the retrieval of the height component is the most precise one as for the same-heading cases, the precision in the north component is the highest. Therefore, in general localization of targets from cross-heading tracks are desirable. The only remaining concern regarding localization using cross-heading tracks is the diameter of the lamp poles which may worsen the accuracy in the east coordinate component. This bias can be estimated and removed if the scatterer is also visible from same-heading tracks which is usually the case.

\begin{figure*}[t!]
\centering
        \includegraphics[width=\textwidth]{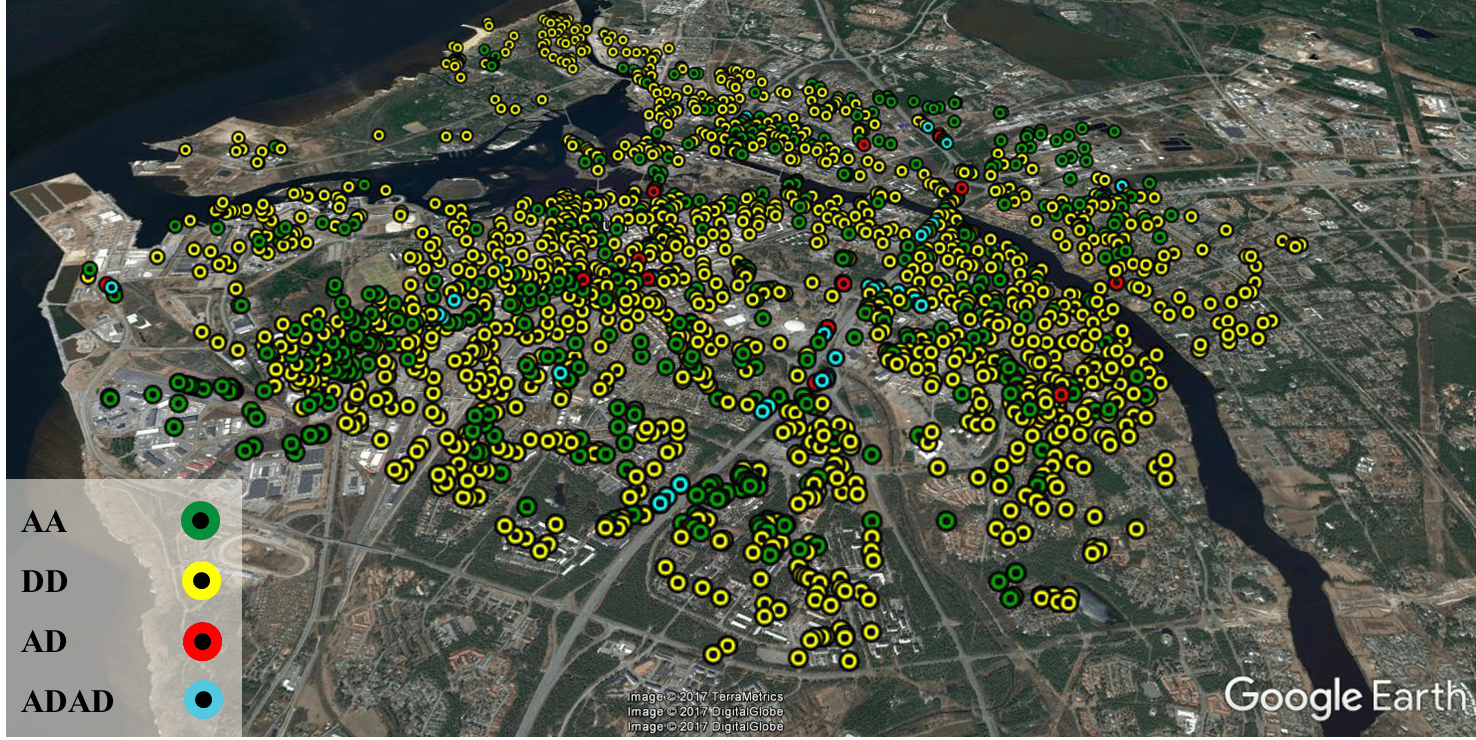}
        \caption{Total number of 2049 GCPs in Oulu color-coded based on the geometry configuration used for their positioning. The underlying optical image is taken from  Google Earth.}\label{fig:GE_Final_Results}
\end{figure*}

\begin{table*}[tbh!]
\centering
\ra{1.3}
\caption{\textsc{Averaged statistics based on the least squares estimated \ac{3D} coordinate standard deviations in Oulu. The letters A and D stand for ascending and descending geometries, respectively. The sample mean and standard deviation are denoted by $\mu$ and $\sigma$, and $S_{[ENH]}$ represent the local coordinates standard deviations within 95$\%$ confidence level}}
  \begin{tabulary}{0.8\textwidth}{|C|CCCCCCC|}
    \hline
    \textbf{Geometry} & \textbf{Nr. Scatterers} & $\mu_{\text{s}_\text{E}}~[\text{cm}]$ & $\mu_{\text{s}_\text{N}}~[\text{cm}]$ & $\mu_{\text{s}_\text{H}}~[\text{cm}]$ & $\sigma_{\text{s}_{\text{E}}}~[\text{cm}]$ & $\sigma_{\text{s}_{\text{N}}}~[\text{cm}]$ & $\sigma_{\text{s}_{\text{H}}}~[\text{cm}]$ \\
  \hline
    AA      & 565  & 17.73 & 5.04 & 15.87  & 11.98  & 2.63 & 11.09 \\

    DD      & 1417 & 15.08 & 3.80 & 16.71  & 10.38 & 2.10 & 11.30  \\

    AD      & 24   & 2.26 & 2.50  & 1.75 & 0.99 & 1.11 & 0.83 \\

    ADAD    & 43   & 1.17 & 1.40  & 1.12 & 0.42 & 0.55 & 0.37 \\ \cline{1-8}
  \end{tabulary}
  \label{tab:Oulu_res}
  \end{table*}

The distribution of the total 2049 generated \ac{gcp}s is visualized on the optical image of Oulu in Google Earth in Fig. \ref{fig:GE_Final_Results}. The scatterers are color-coded based on the underlying geometry configuration used for their localization. It is seen that almost the entire area of Oulu is covered with the generated \ac{gcp}s. The ones from the same-heading geometries cover the built areas while the ones from cross-heading configuration include the base of lamp poles, street lights and traffic lights.

\section{Summary, Conclusion and Outlook}
\label{sec:conc}

In this paper we described a processing chain for automatic detection and positioning of opportunistic \ac{ps}s which are visible from multi-aspect \ac{tsx} images. This paves the way for generation of \ac{gcp}s from \ac{sar} data only.\par

Three algorithms have been recommended for identical PS detection which are different in terms of number of generated \ac{gcp}s and the subsequent positioning precision. The method based on the \ac{psi} fusion algorithm is able to provide point correspondences even on buildings and areas with complex scattering mechanisms in \ac{sar} images. Therefore, a large number of potential \ac{ps} pairs can be obtained which normally cover the entire scene. The downside of the method is that \ac{sar} image stacks are required for which a complete \ac{psi} processing has to be performed separately before being able to find the \ac{ps} correspondences. Furthermore, the method is usually applicable only for same-heading \ac{ps} point clouds. Consequently in terms of localization precision with the subsequent stereo \ac{sar}, the relative error in the cross-range is larger than the error for range and azimuth components. Another disadvantage is that many of the initial \ac{ps} correspondences cannot be considered useful candidates for stereo \ac{sar} and have to be eliminated later in the processing, because the registration is only performed within the limits of the \ac{psi} \ac{3D} localization quality for which \ac{ps} pairs with distances of up to five meters are detected. The detection algorithm based on external optical data is quite straightforward to implement and provides identical scatterers that are visible from cross-heading orbits leading to higher localization precision. The disadvantage of the method is that for reliable detection of lamp poles, the spatial resolution of the optical image should be in the sub-decimeter regime. Moreover, the method is highly prone to detecting other linear structures as shadows of lamp poles and therefore more sophisticated object detection algorithms are recommendable. The method based on vector road network data, similar to the optical method, provides candidates which are suitable to be localized from cross-heading geometries. Also the external data is freely accessible for most locations. The disadvantage of the method is that a co-registration on one master has to be carried out for each stack and the amplitude data must be calibrated.\par

It has been shown that the \ac{gcp} generation processing chain is quite flexible as it allows the user to constraint the number and the quality of the candidate \ac{gcp}s either from the start of the procedure, by selecting different distance thresholds for detection or trimming the data based on estimated phase noise time series, or at the final step of the procedure based on the posterior quality indicators obtained from stereo \ac{sar}.\par

By applying the algorithm to two test sites, it has been demonstrated that it is capable of positioning natural \ac{ps}s with precision values ranging from 2 cm and 4-5 cm, for cross-heading AD and ADAD configurations respectively, to approximately 20 cm for candidates extracted from same-heading geometries. As it was expected, the difference in the viewing geometries of the observed \ac{ps} has the highest impact on the final localization precision followed by number of acquisitions used in stereo \ac{sar} processing, the \ac{scr} of the target and the quality of external error corrections. Furthermore, as a preliminary cross-comparison, the estimated ellipsoidal height of the retrieved candidates in Berlin were compared to the corresponding height of a LiDAR data which reported an average bias of approximately 13 cm.\par

The produced absolute \ac{gcp}s have ample of applications in geodesy and absolute mapping. These points may substitute the conventional \ac{gcp}s that are required for geo-referencing of satellite imagery which are usually surveyed at the field by \ac{gnss}. They can be further integrated as absolute reference points into multi-pass \ac{insar} techniques. Furthermore, they can be used to detect long-term ground motions with small magnitudes and low spatial frequency which are invisible to phase-based \ac{insar} methods.\par
The future work will focus on smart pre-selection of the \ac{ps} candidates by including the information obtained from \ac{pta}, using integrated side-lobe ratio (ISLR), to robustly remove the \ac{ps} candidates which are located too close to each other. Furthermore, the stereo \ac{sar} processing could benefit from weighting the initial timing observations of the \ac{ps}s based on their respective \ac{scr} or \ac{adi} values and also can be carried out with robust parameter estimation schemes like M-estimator. Moreover, it is desirable to carry out \ac{gnss} measurements at selected test sites to be able to correctly validate the absolute accuracy of the generated \ac{gcp}s. Finally, it is important to note that the proposed methodology is tailored to detection and absolute localization of \ac{gcp}s in urban area where a large number of \ac{ps}s are available. In applications where the investigated scene includes mainly non-urban area, it is recommendable to employ artificial \ac{ps}s such as corner reflectors or active transponders.


%

\section*{Acknowledgment}
From the Remote Sensing Technology Institute (IMF) of the DLR, the authors would like to thank Dr. Ulrich Balss for providing the routines for \ac{pta}, Mr. Fernando Rodriguez Gonzalez for his technical support on \ac{psi} processing of Oulu using the PSI-GENESIS and Mr. Nan Ge for re-ordering the TerraSAR-X images of Berlin with updated L1B product files. We are also grateful to Dr. Heiko Hirschm{\"u}ller of the DLR robotics institute for providing us with the optical data of Berlin. The LiDAR  data of Berlin have been provided by Land Berlin (EU EFRE project)  and  Landesamt  f{\"u}r  Vermessung  und  Geoinformation Bayern. The authors gratefully acknowledge the Gauss Centre for Supercomputing e.V. (www.gauss-centre.eu) for funding this project by providing computing time on the GCS Supercomputer SuperMUC at Leibniz Supercomputing Centre (LRZ, www.lrz.de).

\ifCLASSOPTIONcaptionsoff
  \newpage
\fi



\bibliographystyle{IEEEtran}
\bibliography{References}
\end{document}

%% file: acronym.tex
\DeclareAcronym{sar}{
  short = SAR ,
  long  = Synthetic Aperture Radar ,
}

\DeclareAcronym{3D}{
  short = 3-D ,
  long  = three-dimensions ,
}

\DeclareAcronym{2D}{
  short = 2-D ,
  long  = two-dimensions ,
}

\DeclareAcronym{ps}{
  short = PS ,
  long  = Persistent Scatterer ,
}

\DeclareAcronym{gcp}{
  short = GCP ,
  long  = Ground Control Point ,
}

\DeclareAcronym{tsx}{
  short = TS-X ,
  long  = TerraSAR-X ,
}

\DeclareAcronym{gnss}{
  short = GNSS ,
  long  = Global Navigation Satellite System ,
}

\DeclareAcronym{insar}{
  short = InSAR ,
  long  = Interferometric SAR ,
}

\DeclareAcronym{tmsp}{
  short = TMSP ,
  long  = TerraSAR-X multimode SAR processor ,
}

\DeclareAcronym{iers}{
  short = IERS ,
  long  = International Earth Rotation and Reference Systems Service ,
}

\DeclareAcronym{scr}{
  short = SCR ,
  long  = Signal-to-Clutter-Ratio ,
}

\DeclareAcronym{psi}{
  short = PSI ,
  long  = Persistent Scatterer Interferometry ,
}

\DeclareAcronym{utm}{
  short = UTM,
  long  = Universal Transverse Mercator ,
}

\DeclareAcronym{ncc}{
  short = NCC,
  long  = Normalized Cross-Correlation ,
}

\DeclareAcronym{adi}{
  short = ADI,
  long  = Amplitude Dispersion Index ,
}

\DeclareAcronym{pta}{
  short = PTA,
  long  = Point Target Analysis ,
}